\documentclass{article}
\usepackage{graphicx} 
\usepackage{subcaption}
\usepackage{amsmath,amsthm,amsfonts,amssymb,mathrsfs,dsfont,mathtools}
\usepackage[margin=1in]{geometry}
\usepackage{color, xcolor}
\usepackage{framed}
\usepackage{soul}
\usepackage{booktabs}
\usepackage{hyperref}       
\usepackage{url} 
\usepackage{natbib}
\usepackage[noblocks]{authblk}
\usepackage{threeparttable}
\usepackage{comment}

\newcommand{\rc}[1]{{\color{red}{#1}}}

\newcommand{\bc}[1]{{\color{blue}{#1}}}

\newcommand{\lefteq}[1]{%
  \noindent\makebox[\textwidth][l]{%
    $\displaystyle #1$
  }%
}

\newcommand{\fredericcomment}[1]{%
  \vskip 2mm
  \noindent
  \colorbox{blue!10}{%
    \parbox{\dimexpr\linewidth-2\fboxsep\relax}{%
      \textcolor{blue}{\textbf{#1}}\\[1mm]
      \hfill\textcolor{blue}{\textbf{-- Frederic}}%
    }%
  }%
  \vskip 2mm
}

\newcommand{\zaniarcomment}[1]{%
  \vskip 2mm
  \noindent
  \colorbox{magenta!10}{%
    \parbox{\dimexpr\linewidth-2\fboxsep\relax}{%
      \textcolor{magenta}{\textbf{#1}}\\[1mm]
      \hfill\textcolor{magenta}{\textbf{-- Zaniar}}%
    }%
  }%
  \vskip 2mm
}

\begin{document}
\title{Learning to Hedge Swaptions\thanks{Godin is funded by NSERC (RGPIN-2024-04593). The authors report there are no competing interests to declare.}}
\author[a]{Zaniar Ahmadi}
\author[,a,b]{Frédéric Godin\thanks{Corresponding author. \vspace{0.2em} \newline
{\mbox{\hspace{0.47cm}} \it Email addresses:}
\href{mailto:zaniar.ahmadi@concordia.ca}{zaniar.ahmadi@concordia.ca} (Zaniar Ahmadi), \href{mailto:frederic.godin@concordia.ca}{frederic.godin@concordia.ca} (Fr\'ed\'eric Godin)}}

\affil[a]{{\small Concordia University, Department of Mathematics and Statistics, Montr\'eal, Canada}}
\affil[b]{{\small Quantact Laboratory, Centre de Recherches Math\'ematiques, Montr\'eal, Canada}}
\date{\today}
\maketitle

\begin{abstract}
\vspace{-5pt}
This paper investigates the deep hedging framework, based on reinforcement learning (RL), for the dynamic hedging of swaptions, contrasting its performance with traditional sensitivity-based rho-hedging. We design agents under three distinct objective functions---mean squared error, downside risk, and Conditional Value-at-Risk---to capture alternative risk preferences and evaluate how these objectives shape hedging styles.
Relying on a three-factor arbitrage-free dynamic Nelson-Siegel model for our simulation experiments,
our findings show that near-optimal hedging effectiveness is achieved when using two swaps as hedging instruments. 
Deep hedging strategies dynamically adapt the hedging portfolio's exposure to risk factors across states of the market. In our experiments, their out-performance over rho-hedging strategies persists even in the presence some of model misspecification. These results highlight RL’s potential to deliver more efficient and resilient swaption hedging strategies.

\bigskip 



\end{abstract}

\pagebreak
\section{Introduction}

Interest rate derivatives play a central role in modern financial markets, with interest rate swaps and swaptions serving as foundational instruments for risk management and asset-liability transformation. Swaptions, which are options on interest rate swaps, are particularly vital for hedging interest rate volatility and convexity risk, offering flexibility in managing contingent liabilities and exposures. They are widely used by institutions such as insurance companies, pension funds and structured product issuers, and are integral to the pricing and risk management of callable bonds, mortgage-backed securities and other complex instruments. According to the Bank for International Settlements ~\citep{bis2024otc}, the global notional value of outstanding interest rate derivatives exceeds \$540 trillion, with swaptions comprising over \$46 trillion of this total, reflecting their deep integration into the financial ecosystem.

Research on the hedging of swaptions has been ongoing for years. \citet{jamshidian1997libor} demonstrates that a swaption can be replicated using a finite number of zero-coupon bonds, establishing a direct connection between swaptions and fundamental fixed-income contracts. 
\citet{brace2001towards} compute the Greeks of swaptions under the LIBOR market model (LMM) and show that the resulting hedging strategy aligns closely with that obtained with the Black swaption pricing formula.
The effectiveness of delta hedging strategies within LMM is extensively explored by \citet{tim_dun_simulated_2001}. They employ Monte Carlo simulations to show that while at-the-money swaptions can be hedged effectively, significant errors persist for in- and out-of-the-money contracts due to model limitations and sensitivity to volatility and payoff asymmetry. Research highlights the advantages of multi-factor frameworks for the representation of the term structure, which allow interest rates associated with different maturities to move independently and thus to capture the complex shifts, twists, and curvature observed in real markets. \citet{driessen2003performance} find that multi-factor term structure models outperform single-factor models for hedging interest rate derivatives such as caps and swaptions because they capture yield curve movements more accurately. \citet{fan2003hedging} demonstrate that higher-order multifactor Heath-Jarrow-Morton models significantly reduce hedging errors in the swaption market, while \citet{fan2007pricing} further show that at least two factors are required to robustly capture the yield curve dynamics, which is necessary for effective swaption hedging. \citet{an2008compatibility} compare one-factor LIBOR and swap market models, concluding that although they yield comparable results, significant biases and limitations underscore the need for more sophisticated or multi-factor methods. In addition, \citet{cresnikanalysis} evaluates the hedging performance of the \citet{black1976pricing}, \citet{hull1990pricing}, and SABR \citep{hagan2002managing}
models using both historical and synthetically generated scenarios, finding that stochastic volatility and multifactor models such as SABR provide superior risk reduction versus simpler one-factor alternatives. On a discordant note, \cite{pietersz2010comparison} state that hedging performance on swaptions with single-factor models can be satisfactory, with smile effects having more material impacts than rates correlations. Recent research applies machine learning to address the complexity and high dimensionality of swaption hedging. \citet{wang2018deep} introduce a deep learning-based BSDE solver within the LMM framework, utilizing a dedicated neural network architecture that efficiently prices and hedges Bermudan swaptions, computes Greeks, and handles high-dimensional problems. 

Conventional hedging strategies, whether based on static hedges or dynamically updated sensitivities, remain fundamentally local in nature. 
They typically optimize hedging positions one period at a time, without accounting for compounding risk premia or interaction between hedging errors of the future time periods. Deep reinforcement learning (DRL), as in ~\citet{buehler2019deep}, offers an alternative to conventional schemes and allow optimizing policies jointly over multiple periods. The use of DRL approaches for hedging optimization is referred to as deep hedging. The advantages of deep hedging lie in enabling multi-stage optimality, offering flexibility in targeting different objective functions, and providing the ability to account for physical distribution features such as risk premia.  Such an approach can handle high-dimensional environments including multiple realistic characteristics such as transaction costs and market frictions, stochastic volatility \citep{cao2021deep}, and dynamic implied volaility surfaces \citep{franccois2024enhancing,franccois2025deep}. The deep hedging agent learns flexible, state-dependent hedging policies through interaction with simulated or historical market environments. For an overview of methodologies and trends in deep hedging, see the survey by \citet{pickard2023deep}. 

To the best of our knowledge, with the exception of \cite{oya2024deep}, there are currently no other published works that employ deep hedging for swaptions.
This study aims to fill this gap by developing a deep hedging framework for hedging European swaptions when the spot rate dynamics follows a multi-factor short rate process. More specifically, we rely on the discrete-time arbitrage-free Nelson–Siegel model (DTAFNS) of \cite{eghbalzadeh2024discrete}, which is a discrete-time version of the arbitrage-free Nelson–Siegel model of \cite{christensen2011affine}.
This represents an important advancement in tackling the unique challenges posed by interest rate derivatives hedging and in broadening the scope of neural-network-based hedging strategies. A key difference between this paper and that of \cite{oya2024deep} is that they rely on the Swap market Bergomi model, whereas we rely on a parsimonous three-factor model for dynamics of the term structure. The latter choice leads to better interpretability of trading strategies and clearer insight into the source of out-performance of the deep hedging model over benchmarks.

This study highlights multiple advantages of using deep hedging policies for hedging interest rate derivatives. First, we observe general superior performance over rho-hedging, including under some model misspecification. To achieve multi-stage optimality, the deep hedging agent leverages parametric yield-curve dynamics to better manage  exposure to risk factors and associated risk premia. Deep hedging effectively captures higher-order risk exposure and pathwise dynamics (e.g. reflecting previous hedging errors) that traditional approaches fail to address. Moreover, we demonstrate how various risk preferences can be encoded through different choices for the objective functions. In particular, we train agents under mean squared error, downside risk, and Conditional Value-at-Risk (CVaR) objectives, showing how each criterion induces a distinct hedging style. This comparative analysis provides new insights into the trade-offs between hedge effectiveness in regular scenarios, tail-risk protection, and premium harvesting. Additionally, we evaluate the marginal value added of the multiple hedging swaps we consider for inclusion in the hedging portfolio. Our results show that the use of two swaps is sufficient to achieve near-optimal hedging effectiveness. All these aforementioned findings underscore the potential of deep hedging as a robust and economically meaningful framework for interest rate derivatives risk management.

The remainder of this paper is organized as follows. 
Section~\ref{se:MarketEnv} describes the market environment, the DTAFNS model, and the associated pricing of interest rate instruments such as swaps and swaptions. Section~\ref{se:HedgingFramework} introduces the proposed hedging framework, including the construction of the hedging portfolio, the deep hedging approach, and benchmark strategies based on factor sensitivities. Section~\ref{se:ModelSetup} details the experimental design, and neural network architectures employed for both pricing and hedging tasks. Section~\ref{se:ResultAn} presents the empirical results, comparing the performance of deep hedging strategies against traditional benchmarks. Section~\ref{sec:conclusion} concludes.
The code used to produce numerical results of this paper is available on Github.\footnote{See \href{https://github.com/zaniara3/swaption_deep_hedging}{https://github.com/zaniara3/swaption\_deep\_hedging} and \href{https://github.com/zaniara3/swaption_pricing_KAN}{https://github.com/zaniara3/swaption\_pricing\_KAN}.}

\section{Market Environment and Financial Instruments} \label{se:MarketEnv}

We assume that the short-rate dynamics follow the DTAFNS model proposed by \cite{eghbalzadeh2024discrete}. The DTAFNS model offers a tractable, arbitrage-free framework for yield curve modeling in discrete time. Its closed-form solutions and compatibility with financial applications make it a valuable tool for pricing, risk management, and forecasting in stochastic interest rate environments.

The market operates in discrete time with time steps $t = 0, 1, \ldots, T$, each separated by a fixed time increment $\Delta$ (in years). The financial market is modeled on a filtered probability space $(\Omega, \mathcal{F}, \mathbb{P})$, where $\mathbb{P}$ denotes the real-world (physical) probability measure and $\mathcal{F} := \{\mathcal{F}_t\}_{t=0}^T$ represents the natural filtration encoding the information flow in the market. The time-$t$ term structure is driven by three factors, $X_t^{(1)}$, $X_t^{(2)}$ and $X_t^{(3)}$, each of which is $\mathcal{F}_t$-measurable.
The short rate $r_t$ prevailing over the interval $[t, t+1)$ is specified as
\begin{equation} \label{eq:short_rate}
    r_t = X_t^{(1)} + X_t^{(2)}.
\end{equation}
The factors vector $X_t = [ X_t^{(1)}, X_t^{(2)}, X_t^{(3)} ]^\top$ is latent and evolves under measure $\mathbb{P}$ according to
\begin{equation} \label{eq:factor_p_measure}
    X_{t+1} = X_t + \kappa^{\mathbb{P}} (\theta^{\mathbb{P}} - X_t) + \Sigma Z_{t+1}^{\mathbb{P}}.
\end{equation}
Here, $\kappa^{\mathbb{P}} = [\kappa^{\mathbb{P}}_{i,j}] \in \mathbb{R}^{3 \times 3}$ is the mean-reversion parameters matrix, $\theta^{\mathbb{P}} = [ \theta^{\mathbb{P}}_1, \theta^{\mathbb{P}}_2, \theta^{\mathbb{P}}_3 ]^\top$ is the long-term average level of the factors, and $\Sigma =[\Sigma_{i,j}] \in \mathbb{R}^{3 \times 3}$ is a diagonal volatility matrix with strictly positive entries. The innovations $\{Z_t^{\mathbb{P}}\}_{t=1}^T$ are i.i.d. standard multivariate Gaussian vectors under $\mathbb{P}$ with contemporaneous correlations
\[
\rho_{ij}^{(Z)} := \mathrm{corr}(Z_{t,i}^{\mathbb{P}}, Z_{t,j}^{\mathbb{P}}), \quad \text{for } i, j = 1,2,3.
\]
The initial factor value $X_0 = x_0 \in \mathbb{R}^3$ is fixed and is assumed to be known.

To enable arbitrage-free pricing of interest rate derivatives such as bonds, bond options, swaps, and swaptions, \cite{eghbalzadeh2024discrete} specify risk-neutral dynamics under an equivalent martingale measure $\mathbb{Q} \sim \mathbb{P}$. The factor dynamics under $\mathbb{Q}$ is assumed to be of the form
\begin{equation} \label{eq:factor_q}
    X_{t+1} = X_t + \kappa^{\mathbb{Q}}(\theta^{\mathbb{Q}} - X_t) + \Sigma Z_{t+1}^{\mathbb{Q}},
\end{equation}
where  $\kappa^{\mathbb{Q}}$ and $\theta^{\mathbb{Q}}$ are obtained by solving the system of equations
 \begin{equation}
\left\{
\begin{aligned}
    \kappa^{\mathbb{Q}} &= \kappa^{\mathbb{P}} - \Sigma \gamma, \\
    \kappa^{\mathbb{Q}} \theta^{\mathbb{Q}} &= \kappa^{\mathbb{P}} \theta^{\mathbb{P}},
\end{aligned}
\right.
\end{equation}
for some constant vector $\gamma$, with $\{Z^{\mathbb{Q}}_t\}$ being independent standard Gaussian vectors with correlation matrix $[\rho_{ij}^{(Z)}]^3_{i,j=1}$ under $\mathbb{Q}$.
This dynamics retains the same autoregressive structure as under the physical measure. This consistency is desirable because it preserves analytical tractability and facilitates simulation under both real-world and risk-neutral measures.
Furthermore, it is assumed that matrix $\kappa^{\mathbb{Q}}$ is of the form
\[
\kappa^{\mathbb{Q}} =
\begin{bmatrix}
0 & 0 & 0 \\
0 & \lambda & -\lambda \\
0 & 0 & \lambda
\end{bmatrix},
\quad \text{with } \lambda \in (0,1).
\]
The null first row of $\kappa^{\mathbb{Q}}$ entails that the first factor is non-stationary under $\mathbb{Q}$ despite being stationary under $\mathbb{P}$.

\paragraph{Zero Coupon Pricing.}
The DTAFNS model admits an exponential-affine solution for zero-coupon bond prices. Specifically, for a bond maturing at time $T$, the arbitrage-free price at time $t < T$ is given by
\begin{equation} \label{eq:zc_price_Q}
    P(t, T) = \mathbb{E}^{\mathbb{Q}} \left[ \exp\left( -\Delta \sum_{j=t}^{T-1} r_j \right) \middle| \mathcal{F}_t \right] 
    = A_\tau \exp\left[ -\Delta B_\tau^\top X_t \right],
\end{equation}
where $\tau = T - t$ denotes the number of time steps remaining before maturity, and $B_\tau = [ B_\tau^{(1)}, B_\tau^{(2)}, B_\tau^{(3)} ]^\top$ is a vector of loadings associated with each factor. The mapping $A_\tau$ reflects the yield curve shape when term structure factors are null. Closed-form formulas for $(A_\tau, B_\tau)$ are provided in Appendix \ref{app:zero_coupon_price}.

\paragraph{Interest Rate Swaps Pricing.}
According to \cite{brigo2001interest}, a Payer Forward-Starting Interest Rate Swap (PFS) is a contract with a future effective date $T_\alpha$, 
where one party agrees to pay a fixed interest rate $K$ on a nominal amount $N$ and receive a floating rate over a sequence of dates $\{T_{\alpha+1}, \ldots, T_\beta\}$, where $T_{i+1}=T_i+1$ for all $i$ (with an elapse of $\Delta$ years occurring between $T_i$ and $T_{i+1}$). 
At each payment date $T_i$ in this schedule, the fixed leg pays
\[
N K \Delta,
\]
while the floating leg pays
\[
N L(T_{i-1}, T_i) \Delta,
\]
where $L(T_{i-1}, T_i)$ denotes the simply-compounded spot rate set at $T_{i-1}$ for the period $[T_{i-1}, T_i]$:
\begin{equation*}
    L(T_{i-1},T_i) = \frac{1- P(T_{i-1},T_i)}{ P(T_{i-1},T_i) \Delta}.
\end{equation*} 
The floating leg resets at dates $T_\alpha, T_{\alpha+1}, \ldots, T_{\beta-1}$ and pays at dates $T_{\alpha+1}, \ldots, T_\beta$. Let $\mathcal{T} := \{T_\alpha, \ldots, T_\beta\}$ denote the full tenor schedule of the swap.

The discounted value of the PFS cash flows at a time $t < T_\alpha$ is given by
\[
\sum_{i=\alpha+1}^{\beta} D(t, T_i) N \Delta \left( L(T_{i-1}, T_i) - K \right),
\]
where  $D(t, T_i) = \exp \left( -\Delta\sum_{s=t}^{T_i-1} r_s \right)$  is the stochastic discount factor between time $t$ and the payment date $T_i$.
Using no-arbitrage arguments and risk-neutral pricing, the forward-start swap value can be expressed in terms of observable market instrument prices as
\begin{equation} \label{eq:pfs_price}
    \begin{split}
        \text{PFS}(t, \mathcal{T}, N, K) &= N \Delta \sum_{i=\alpha+1}^{\beta}  P(t, T_i) \left( F(t; T_{i-1}, T_i) - K \right) \\
        &= N \left( P(t, T_\alpha) - P(t, T_\beta) \right) - N K \Delta \sum_{i=\alpha+1}^{\beta}  P(t, T_i),
    \end{split}
\end{equation}
where $F(t; T_{i-1}, T_i)$ denotes the simply-compounded forward rate for the period $[T_{i-1}, T_i]$ 
\[
F(t; T_{i-1}, T_i) = \frac{1}{\Delta} \Big( \frac{P(t,T_{i-1})}{P(t,T_i)} -1 \Big).
\]

The forward swap rate, denoted $S_{\alpha,\beta}(t)$, is the fixed rate that equates the value of the PFS to zero (i.e., the par rate). Solving for this value yields
\[
S_{\alpha,\beta}(t) = \frac{P(t, T_\alpha) - P(t, T_\beta)}{\Delta \sum_{i=\alpha+1}^{\beta} P(t, T_i)}.
\]
This rate is commonly used as a benchmark in swaption pricing and hedging applications.

\paragraph{Swaption Pricing.}

A \textit{European payer swaption} is a financial derivative that grants its holder the right, but not the obligation, to enter into a payer interest rate swap (IRS) at a predetermined future time, typically coinciding with the first reset date of the underlying swap, denoted $T_\alpha$. The length of the underlying swap, given by $T_\beta - T_\alpha$, is referred to as the \textit{tenor} of the swaption.

From \eqref{eq:pfs_price}, the value of the underlying payer IRS at the swaption maturity $T_\alpha$ is
\[
N \Delta  \sum_{i=\alpha+1}^{\beta} P(T_\alpha, T_i) \left( F(T_\alpha; T_{i-1}, T_i) - K \right).
\]
The swaption is exercised only if this value is positive. Thus, the discounted payoff of the payer swaption at a time $t < T_\alpha$ is
\[
ND(t, T_\alpha) \Delta \left( \sum_{i=\alpha+1}^{\beta} P(T_\alpha, T_i) \left( F(T_\alpha; T_{i-1}, T_i) - K \right) \right)^+.
\]
The swaption is said to be \textit{at-the-money} (ATM) if the strike equals the forward swap rate, i.e.
\[
K = K_{\text{ATM}} := S_{\alpha,\beta}(0) = \frac{P(0, T_\alpha) - P(0, T_\beta)}{\Delta \sum_{i=\alpha+1}^\beta  P(0, T_i)}.
\]
It is \textit{in-the-money} (ITM) for the payer swaption if $K < K_{\text{ATM}}$ and \textit{out-of-the-money} (OTM) if $K > K_{\text{ATM}}$. 

Because the holder of the swaption can, when exercising at maturity $T_\alpha$, enter---at zero cost---an ATM payer swap whose fixed rate is the swap rate, the risk-neutral measure can be used to represent the time-$t$ price of a European payer swaption with notional $N$, strike $K$, and tenor structure $\mathcal{T}$ through
\begin{equation*}
    \begin{split}
        \mathrm{PS}\left[t, \mathcal{T}, K, N \right] &= \mathbb{E}^{\mathbb{Q}} \left[ \left. D(t, T_\alpha) \left( N \left( S_{\alpha,\beta}(T_\alpha) - K \right)^+ \Delta\sum_{i=\alpha+1}^{\beta} P(T_\alpha, T_i) \right) \right| \mathcal{F}_t \right] = \mathbb{E}^{\mathbb{Q}} \left[D(t, T_\alpha) \mathcal{P}_{T_\alpha} | \mathcal{F}_t \right]
    \end{split}
\end{equation*}
where $\mathcal{P}_{T_\alpha}$ is the swaption payoff at maturity 
\[
\mathcal{P}_{T_\alpha} = N \left( 1 - P(T_\alpha, T_\beta) - K \Delta \sum_{i=\alpha+1}^\beta  P(T_\alpha, T_i) \right)^+. \ 
\]
As explained in \cite{godin2023pricing}, the price can also be expressed by using the $T_\alpha$-forward measure $\mathbb{Q}^{T_\alpha}$, which uses $P(t, T_\alpha)$ as the numéraire to simplify the pricing formula
\begin{align} \label{eq:swaption_price_T_measure}
\mathrm{PS}\left[t, \mathcal{T}, K, N \right] 
&= P(t, T_\alpha) \mathbb{E}^{\mathbb{Q}^{T_\alpha}} \left[\mathcal{P}_{T_\alpha}| \mathcal{F}_t \right].
\end{align}

This formulation reduces the pricing problem to computing the conditional expectation of a function of future bond prices. Since zero-coupon bond prices under the DTAFNS model are exponential-affine in the factors, the time-$T_\alpha$ bond prices involved in \eqref{eq:swaption_price_T_measure} can be fully characterized by the term structure factors $X_{T_\alpha}$. For the dynamics of the factors under the $T$-forward measure, refer to Appendix \ref{app:forward_measure}.

\section{Hedging Framework and Replicating Portfolio Construction} \label{se:HedgingFramework}

We consider a discrete-time hedging framework where hedging portfolio rebalancing dates are $\{0, 1, \dots, T_\alpha\}$.
The objective is to construct a hedging portfolio to hedge a short position on a European swaption. Without loss of generality, a payer swaption is considered.

To hedge a short position in this option, an agent constructs a dynamic self-financing portfolio comprising $M$ forward-starting payer swaps (which may include the underlying swap) and a money market account that accrues interest at the prevailing short rate. The value of the hedging portfolio at time $t \leq T_\alpha$ is given by
\begin{equation} \label{eq:replicated_port}
    V_{t} = \phi_t^\top  \mathrm{PFS}_{t} + \psi_t e^{r_{t-1} \Delta},
\end{equation}
where $\mathrm{PFS}_t = [\mathrm{PFS}_t^{(1)}, \dots, \mathrm{PFS}_t^{(M)} ]^\top$ is the time-$t$ value of the hedging forward-starting payer swaps, $\psi_t$ is the cash amount held at time $t-1$, which earns interest compounded at rate $r_{t-1}$, and 
the hedging strategy is represented by the predictable process $\{(\phi_t, \psi_t) \}_{t=1}^{T_\alpha}$, where $\phi_t = [\phi_t^{(1)}, \dots, \phi_t^{(M)}]^\top$ is the vector of hedge ratios (i.e., multiplier of the notional value) of the swap positions held at time $t-1$.\footnote{In the case $N=1$ considered in this work, $\phi_t$ represents notional values.}
Under the self-financing condition, the cash holding is determined endogenously as\footnote{In unreported experiments, we have tested the inclusion of transaction costs, which impact the portfolio value. However, the impact of such costs under realistic cost levels reflective of market conditions was very minor, and thus we chose not to include them in the paper.}
\begin{equation}
    \psi_{t+1} = V_{t} - \phi_{t+1}^\top  \mathrm{PFS}_{t}.
\end{equation}

The agent aims to determine an optimal dynamic hedging strategy $\{\phi_t\}^{T_\alpha}_{t=1}$, with the goal of minimizing the hedging error risk at the swaption maturity $T_\alpha$.

We refer to the set $\mathcal{X} = \{ X_t,\, t=0, \dots, T_\alpha \}$ as the path of yield curve factors up to time $T_\alpha$. At each time $t$, the hedger observes the vector $Y_t$, which represents the information taken into account to determine their hedging positions.
The composition of $Y_t$ in our setting is detailed subsequently. Based on this input, the hedge ratios are determined by a policy function $\pi_\theta: Y_{t} \mapsto \phi_{t+1}$, parameterized by a neural network with weights $\theta$:
\begin{equation}
    \label{policyeq}
    \phi_{t+1} = \pi_\theta(Y_{t}).
\end{equation}
This approximation allows representing the sequence of hedging decisions as a parameterized control policy that is learned from simulated market scenarios. 

The terminal hedging error is defined as the difference between the swaption payoff and the final hedging portfolio value:
\[
h_{T_\alpha}(\pi_\theta) = \mathcal{P}_{T_\alpha} - V_{T_\alpha}(\pi_\theta),
\]
where the dependence of the terminal hedging portfolio value $V_{T_\alpha}$ on policy parameters $\theta$ is made explicit.
A positive value of $h_{T_\alpha}$ indicates a hedging shortfall, i.e., the hedging strategy failed to fully offset the liability. The hedger’s objective is to find a policy $\pi_\theta$ that minimizes a risk measure applied to the hedging error:
\begin{equation} \label{eq:objective_fun}
    \arg\min_{\theta} \; \mathcal{J} (\theta), \quad \mathcal{J} (\theta) \equiv \zeta\left( h_{T_\alpha}(\pi_\theta) \right)
\end{equation}
for a given risk measure $\zeta$.
This formulation is closely related to the deep hedging framework introduced in \cite{buehler2019deep}, where trading strategies are optimized end-to-end using neural networks and Monte Carlo simulations. 

\subsection{Deep Hedging Using Reinforcement Learning}\label{sec:DeepHedgingRL}

To assess the performance of a given policy, we use Monte Carlo simulation. We generate $n$ independent paths $\mathcal{X}^{(i)}$, $i = 1, \dots, n$, where each path is a realization of the factor process $\{X_t\}_{t = 1, \dots, T_\alpha}$. For each path and for a fixed parameter vector $\theta$,
we compute the replicating portfolio values using \eqref{eq:replicated_port}, with hedge ratios at each time step being given by \eqref{policyeq}.

Accordingly, we define the terminal hedging error $h_{T_\alpha}$ as a function of both the market path $\mathcal{X}^{(i)}$ and the hedging strategy generated by the policy $\pi_{\theta}$, which we denote by
\[
h_{T_\alpha}^{(i)} := h_{T_\alpha}(\pi_{\theta}; \mathcal{X}^{(i)}).
\]
Estimates of the objective function \eqref{eq:objective_fun} are obtained from the empirical distribution of hedging errors. Denote by $n_b$ the batch size used in training.
The mean squared error (MSE) objective is estimated as
\begin{equation}
\label{eq:mse_objective}
    \hat{\mathcal{J}}_{\text{MSE}}(\theta) = \frac{1}{n_b} \sum_{i=1}^{n_b} \left[ h_{T_\alpha}^{(i)} \right]^2.
\end{equation}
The downside risk (DR) objective penalizes only under-hedging scenarios, and its estimate is
\begin{equation}
\label{eq:dr_objective}
    \hat{\mathcal{J}}_{\text{DR}}(\theta) = \frac{1}{n_b} \sum_{i=1}^{n_b} \left[ \max (  h_{T_\alpha}^{(i)},0)  \right]^2.
\end{equation}
When the objective is to control tail risk, we consider the CVaR as the performance criterion. Following \cite{rockafellar2000optimization}, we define the empirical $\mathfrak{a}$-value-at-risk of the hedging error distribution as
\begin{equation}
\label{eq:empirical_var}
    \hat{v}_\mathfrak{a}^{\,n_b} := \inf \left\{ z \in \mathbb{R} \;\middle|\; \frac{1}{n_b} \sum_{k=1}^{n_b} \mathbf{1}_{\{ h_{T_\alpha}^{(i)} \leq z \}} \geq \mathfrak{a} \right\}.
\end{equation}
The corresponding CVaR$_\mathfrak{a}$ is estimated via:
\begin{equation}
\label{eq:empirical_cvar}
    \hat{\mathcal{J}}_{\text{CVaR}}(\theta) = \hat{v}_\mathfrak{a}^{\,n_b} + \frac{1}{n_b(1 - \mathfrak{a})} \sum_{i=1}^{n_b} \left[ h_{T_\alpha}^{(i)} - \hat{v}_\alpha^{\,n_b} \right]^+,
\end{equation}
This loss encourages the agent to minimize the average excess of large hedging errors over the value-at-risk, aiming to mitigate the severity of highly adverse scenarios.

The optimization problem \eqref{eq:objective_fun} is tackled using mini-batch stochastic gradient descent (SGD). The parameter vector $\theta$ is iteratively updated using:
\begin{equation}
\label{eq:update_theta}
    \theta^{(j+1)} = \theta^{(j)} - \mathfrak{n}_j \nabla_{\theta} \hat{\mathcal{J}}(\theta^{(j)}),
\end{equation}
where $\mathfrak{n}_j$ is the learning rate for iteration $j$, and the gradient $\nabla_{\theta} \hat{\mathcal{J}}(\theta^{(j)})$ is computed using automatic differentiation in \texttt{PyTorch}. 
Here, the superscript $j$ denotes the iteration index of the SGD procedure. At each iteration, a new batch of Monte Carlo paths $\mathcal{X}$ is generated under the current policy $\theta^{(j)}$, and the risk-based objective $\hat{\mathcal{J}}(\theta^{(j)})$ is estimated from this batch. This procedure ensures that parameter updates are always based on fresh simulations reflecting the evolving policy.

\subsection{The Rho-Hedging Benchmark}

In this section, we introduce the rho-hedging strategy that serves as a benchmark for the learning-based methods developed in this paper.

To hedge the swaption position against market risk,  the common rho-hedging procedure neutralizes its sensitivity to the key term structure factors $X_t^{(1)}$, $X_t^{(2)}$, or $X_t^{(3)}$ that drive the evolution of interest rates under the DTAFNS model. This is achieved by equating the factor-wise sensitivities (i.e., the rhos) of the swaption and these of the replicating portfolio. 

When using a single forward-start swap ($M = 1$) in the replicating portfolio, to neutralize the sensitivity to a single factor $k \in \{1,2,3\}$; the hedge ratio must be set to
\[
\phi_{t+1}^{(1)} = \frac{\partial \mathrm{PS}_{t}}{\partial X_{t}^{(k)}} \bigg/ \frac{\partial \mathrm{PFS}^{(1)}_{t}}{\partial X_{t}^{(k)}}.
\]
where $\mathrm{PS}_t$ denotes the time-$t$ value of the payer swaption. 

The choice of the risk factor that is neutralized depends on which risk dimension (e.g., level, slope, or curvature) is deemed most significant for the swaption exposure. As the factor $X_{t-1}^{(k)}$ evolves over time, $\phi_t^{(1)}$ is dynamically updated, resulting in a rho-hedging strategy that rebalances the portfolio to attempt tracking the swaption’s value.

In the case where two swaps are used ($M = 2$), the hedge ratios that neutralize the sensitivity of two of the factors $k \neq j \in \{ 1,2,3\}$ to be chosen are determined by solving the linear system
\[
\underbrace{\begin{bmatrix} 
\frac{\partial \mathrm{PFS}^{(1)}_{t}}{\partial X_{t}^{(j)}} & \frac{\partial \mathrm{PFS}^{(2)}_{t}}{\partial X_{t}^{(j)}} \\ 
\frac{\partial \mathrm{PFS}^{(1)}_{t}}{\partial X_{t}^{(k)}} & \frac{\partial \mathrm{PFS}^{(2)}_{t}}{\partial X_{t}^{(k)}} 
\end{bmatrix}}_{=:Q_t}
\begin{bmatrix} 
\phi_{t+1}^{(1)} \\ \phi_{t+1}^{(2)} 
\end{bmatrix}
=
\underbrace{\begin{bmatrix} 
\frac{\partial \mathrm{PS}_{t}}{\partial X_{t}^{(j)}} \\ 
\frac{\partial \mathrm{PS}_{t}}{\partial X_{t}^{(k)}} 
\end{bmatrix}}_{=:b_t}.
\]
This ensures that the portfolio has a null local sensititivy with respect to both factors $j$ and $k$. This system of linear equations can be generalized for neutralizing three factors using three swaps. 

However, instead of applying above formulas providing exact solutions, we instead apply a regularized least-squares approach to minimize net sensitivites of the portfolio. This enhances numerical stability and incorporates additional sensitivity control.
Specifically, the hedge ratios are obtained from
\[
\phi_{t+1}
= \underset{\phi}{\arg\min} \, \|Q_t\phi - b_t\|^{2}
+ l_1 \|\phi\|^{2}
+ l_2 \|\phi - \phi_{t}\|^{2},
\]
where $l_1$ and $l_2$ are regularization parameters and $Q_t$ and $b_t$ are respectively the matrix and the vector containing sensitivities of all hedging swaps and of the swaption to the various factors. This approach removes abrupt fluctuations in hedging positions that were otherwise (very rarely) observed in our simulations, leading to instability and poor performance. In the vast majority of cases the impact of the regularization on hedging positions is almost null for the mild values parameters $l_1$ and $l_2$ considered in our work.\footnote{In this study we set $l_1=l_2=0.01$.}

Once the hedge ratios are determined, we impose two additional leverage constraints on the outputs which are detailed in Section \ref{se:dynamic_lev_bound}.

The price of a PFS is given by \eqref{eq:pfs_price}, and its gradient with respect to  factor $k = 1,2,3$ is
\begin{equation}
    \begin{split}
        \frac{\partial \mathrm{PFS}_t}{\partial X_t^{(k)}} &= - N \Delta \Big(  
        B_{T_{\alpha}-t}^{(k)} P(t,T_{\alpha}) - 
        B_{T_{\beta}-t}^{(k)} P(t,T_{\beta}) - 
         K \Delta \sum_{i= \alpha+1}^{\beta}  B_{T_{i}-t}^{(k)} P(t, T_i) 
        \Big).
    \end{split}
\end{equation}

\section{Market and Model Setup} \label{se:ModelSetup}
We use the calibrated DTAFNS model proposed by \cite{eghbalzadeh2024discrete} on end-of-month Canadian spot rate curves from January 1986 to January 2022 (434 months) to simulate the yield curve term structure factors. The calibrated parameters are given in Appendix \ref{app:market_data}. 

Based on this model, we investigate the dynamic hedging of a European payer swaption written on a plain-vanilla forward-starting interest rate swap. Specifically, we consider an ATM swaption with a maturity of $T_\alpha =60$ months (5 years with $\Delta = 1/12$), giving the holder the right to enter into a 5y$\times$10y payer swap, i.e. a swap that starts in five years and runs for ten years (a five-years-forward, ten-year-tenor swap), with a fixed rate of $K =2.5083$\% and unit notional ($N = 1$). The hedging portfolio is rebalanced monthly, using forward-starting payer swaps as hedging instruments. Hedging swaps include the underlying swap and possibly other swaps with different tenors or maturities. 
The swaption price at time $t = 0$ serves as the initial value $V_0$ of the replicating portfolio.

\subsection{Hedging Network Architecture}

For the hedging agent, we employ a four-layer fully connected neural network (FCNN) to approximate the mapping from observed market features to hedge ratios. The width of the respective FCNN hidden layers is $[8, 32, 32, 8]$, followed by a final output layer of size $M$, where $M$ is the number of forward-starting swaps used in the replicating portfolio. For some input $Y_t$, the network output is a vector $\phi_{t+1} = \pi_\theta(Y_t)$ representing the hedge ratios for the $M$ swaps at time $t$ before any constraint is imposed. Once the unconstrained hedge ratios have been determined, we impose two constraints on the outputs to reflect risk limits analogous to these applied in practice. The details about the constraints are provided in Section \ref{se:dynamic_lev_bound}. 

The input vector $Y_t$ which includes the current values of the term structure factors $X_t$, the time-to-maturity $(T_\alpha - t) \Delta$, and the current portfolio value $V_t$.
All inputs except time-to-maturity are first normalized for each batch. The processed input then passes through a sequence of fully connected layers, each followed by a Mish activation function \citep{misra2019mish}
\[
x \mapsto x \cdot \tanh(\ln(1 + e^x)),
\]
which provides smooth and non-monotonic transformations.
No dropout is applied, and the network is kept relatively shallow to avoid overfitting given the structured nature of the inputs. We employ a linear activation function for the final layer to determine the positions in the hedging swaps.

The neural network parameters are initialized using the Kaiming uniform method \citep{he2015delving}, which defines the initial parameter vector $\theta^{(0)}$. It is designed for networks with nonlinear activations and helps maintain stable gradients during training.

Training is performed using the Adam optimizer from \cite{kingma2014adam} with an initial learning rate of $5 \times 10^{-3}$. 
The model is trained for a maximum of 800 epochs. The learning rate is reduced using a plateau-based scheduler, and early stopping is applied with a patience threshold of 200 epochs to avoid overfitting. Each epoch consists of multiple mini-batches of size $n_b=2048$, drawn from a simulated dataset of $n=100{,}000$ Monte Carlo paths. The training objective is to minimize a user-specified risk measure, such as MSE, DR, or CVaR as defined in Section \ref{sec:DeepHedgingRL}, applied to the terminal hedging error.

\subsection{Leverage Bounds} \label{se:dynamic_lev_bound}

The hedging positions outputted by the neural network and rho-hedging positions are unconstrained. In practice, trading desks face binding limits: treasury funding and credit lines limits, clearing margin and liquidity haircut constraints, counterparty/concentration limits, and book-level risk controls such as gross notional and DV01/PV01 caps.\footnote{%
DV01 (Dollar Value of a Basis Point) measures first-order price sensitivity to a 1 bp parallel shift in the yield curve.
PV01 (Present Value of 1 bp) is the present value change from a 1 bp change in the fixed rate of a swap.}
To reflect these operational realities and to avoid pro-cyclical “lever up on gains, delever on losses” behavior,\footnote{%
A constraint scaled by the current hedging portfolio value $|V_t|$ expands risk capacity after gains and contracts it after losses, inducing buy-high/sell-low rebalancing and elevated turnover. We mitigate this by using a stabilized basis $(|V_t|+B)$ in our limits.}
we impose two complementary constraints on our notional positions
\begin{align}
\text{(per-leg exposure)}\quad & |\phi_{i,t}|\,|\mathrm{PFS}_{i,t}| \;\le\; L_{\text{per}}\,(|V_t| + B), \quad \forall i,\\
\text{(gross exposure)}\quad & \sum_{i=1}^K |\phi_{i,t}|\,|\mathrm{PFS}_{i,t}| \;\le\; L_{\text{gt}}\,(|V_t| + B).
\end{align}

The absolute value of the $\phi_{i,t}$, the hedging notional position for leg $i$, and that of the forward-starting swap price $\mathrm{PFS}_{i,t}$, serve as the exposure factor in dollars, consistent with desk practices for notional caps. The first constraint prevents any single leg from dominating the book (mitigating ill-conditioning and concentration risk), while the second controls overall leverage, in line with standard gross exposure or DV01 guardrails at the portfolio level. The scaling term $(|V_t|+B)$ provides a stabilized risk budget: $V_t$ denotes the current portfolio value (taken in absolute value in implementation), and $B\!\ge\!0$ is a fixed liquidity/capital buffer that (i) avoids vanishing limits when $|V_t|$ is small, and (ii) curbs pro-cyclical upsizing when $|V_t|$ spikes. The leverage multipliers $L_{\text{per}}$ and $L_{\text{gt}}$ are policy parameters reflecting the desk’s risk appetite and external constraints (funding, margin, and capital).\footnote{By setting $V_t = V_0$, one can change the dynamic leverage to a static leverage constraint.}
These bounds keep the hedge implementable across liquidity regimes, 
align with real-world risk controls, and improve numerical stability during learning and backtesting.

Let the (unconstrained) per-leg dollar exposures be
\[
e_{i,t} \;=\; \lvert \varphi_{i,t}\rvert \,\lvert \mathrm{PFS}_{i,t}\rvert,\qquad i=1,\dots,M,
\]
where $\varphi_{i,t}$ is the unconstrained position on swap $i$ at time $t$ obtained either from  the neural network for deep hedging, or from regularized least-squares for rho-hedging.
Limits scale with a stabilized basis
\[
B_t \;=\; \lvert V_t\rvert + B,
\]
We enforce a per-leg exposure cap and a gross (portfolio-level) exposure cap
\[
0 \le x_{i,t} \le L_{\mathrm{per}}\,B_t,
\qquad
\sum_{i=1}^M x_{i,t} \le L_{\mathrm{gt}}\,B_t,
\]
where $x_{i,t} = |\phi_{i,t}||\mathrm{PFS}_{i,t}|$.
At each $t$, we project $e_t=[e_{1,t},\dots,e_{M,t}]$ onto the feasible set by
\[
x_t^\star
=
\arg\min_{x\in\mathbb{R}^M}\;\big\|x-e_t\big\|_2^2
\quad\text{s.t.}\quad
0\le x_i\le L_{\mathrm{per}}\,B_t\ (i=1,\dots,M),\;
\sum_{i=1}^M x_i\le L_{\mathrm{gt}}\,B_t.
\]
Feasible hedge notionals are then recovered, preserving direction,
\[
\phi_{i,t} \;=\; \operatorname{sign}\!\big(\varphi_{i,t}\big)\;\frac{x_{i,t}^\star}{\lvert \mathrm{PFS}_{i,t}\rvert}\,.
\]

We employ this projection on both benchmark and RL weights to obtain a fair comparison\footnote{We use $L_{per} = 2$, $L_{gt} = 3$ and $B = 1$.}. Furthermore, such constraints are imposed post-hoc instead of being integrated during the training of the policy since these constraints reduce the smoothness of the objective function and lead to less stable training.

\subsection{Swaption prices and their derivatives} \label{se:SWP}
Swaption prices provided by \eqref{eq:swaption_price_T_measure} are not available in closed-form. To avoid recomputing estimates of such prices through Monte Carlo simulation every time a price is needed, we pre-compute prices of swaptions offline with a Kolmogorov-Arnold neural network. The methodology is explained in Appendix~\ref{sec:SwaptionPricingMetho}. This approach is also used to compute the first-order sensitivity of swaption prices with respect to term structure risk factors, as also detailed in that section.

\section{Numerical Experiments} \label{se:ResultAn}

In this section, we evaluate the performance of the deep hedging strategy and compare it to that of conventional rho-hedging approaches. The evaluation is performed on $N^{(OOS)}=100,\!000$ out-of-sample paths generated with the same parameters that are used to generate training set paths. 

\subsection{Hedging Performance Metrics}

We assess the distribution of terminal hedging errors using a comprehensive set of risk and performance metrics that are hereby provided.

\paragraph{Mean.}
The \emph{mean} of the hedging errors $h_{T_\alpha}^{(i)}, i =1,\ldots, N^{(OOS)}$ captures the average hedging bias, indicating whether a strategy systematically \emph{overhedges} (negative mean) or \emph{underhedges} (positive mean).

\paragraph{Root Mean Squared Error (RMSE).}
The RMSE, which is calculated as the square root of the MSE obtained by applying formula \eqref{eq:mse_objective} to out-of-sample paths, penalizes both over- and underhedging symmetrically.

\paragraph{Root Downside Risk (RDR).}
Computed through the square root of \eqref{eq:dr_objective} applied on out-of-sample paths, the root downside risk metric quantifies the impact of positive hedging errors, assessing the severity of underhedging losses.

\paragraph{Tail Risk.}
Tail risk is measured with CVaR$_{99\%}$, which gives the average value of the worst 1\% hedging losses $h_{T_\alpha}^{(i)}$.

\paragraph{Probability of underhedging.} We also compute the \emph{probability of underhedging}
\[
P(HE > 0) = \frac{1}{N^{(OOS)}} \sum_{i=1}^{N^{(OOS)}} \mathds{1}_ { \{h_{T_\alpha}^{(i)} > 0 \} },
\]
which reflects the frequency at which the hedging strategy fails to fully replicate the swaption payoff.

\paragraph{Hedging Risk Reduction (HRR).}
To evaluate the effectiveness of the hedging strategy relative to an unhedged position, we compute the \emph{hedging risk reduction}, defined as
\[
\text{HRR} = 1 - \frac{\text{Std}(h_{T_\alpha})}{\text{Std}(h^{\text{unhedged}}_{T_\alpha})},
\]
where $h^{\text{unhedged}}_{T_\alpha}$ denotes the hedging error in absence of hedging, i.e. when $\phi \equiv \mathbf{0}$. When no hedging is applied, the swaption premium received  at time $0$ is deposited in a bank account and accrues risk-free interest until the swaption maturity. 
The HRR metric represents the percentage reduction in hedging error volatility achieved by hedging.

\paragraph{Trading Intensity (TI).}
TI measures the extent of replicating portfolio position fluctuations between consecutive time steps, across all hedging instruments.
\[
\text{TI} = \frac{1}{N^{(OOS)}} \sum_{i=1}^{N^{(OOS)}} \sum_{t=1}^{T_\alpha} \sum_{k=1}^M \left| \phi_t^{(k)} - \phi_{t-1}^{(k)} \right|.
\]

\paragraph{Dynamic Tracking Error (DTE).}
Dynamic Tracking Error measures the pathwise deviation between the hedged portfolio value and the swaption price across the hedging horizon. For each path \( i \), it is defined as
\[
\text{DTE}_i = \sqrt{\frac{1}{T_\alpha} \sum_{t=1}^{T_\alpha} \big( V_{i,t} - \mathrm{PS}_{i,t} \big)^2},
\]
where \( \mathrm{PS}_{i,t} \) is the swaption price at time \( t \) for path \( i \).
The average DTE across all paths is
\[
\text{DTE} = \frac{1}{N^{(OOS)}} \sum_{i=1}^{N^{(OOS)}} \text{DTE}_i.
\]



\subsection{Hedging Performance}

This section presents an assessment of the performance of our proposed RL method through Monte Carlo simulation. The RL method is compared to conventional dynamic rho-hedging approaches. The impact of the number of hedging instruments on hedging performance is assessed, with either one, two or three swaps being included in the hedging portfolio.
The single hedging swap case employs the underlying forward-starting swap, which has a start date on the swaption's maturity and whose tenor is the same as that of the swaption. For the two-swap scenario, we augment the hedging portfolio by including an additional 10y$\times$2y swap, i.e. two-year-tenor swap starting in ten years. This second swap is chosen to load primarily on the long end of the curve, capturing movements in long-term rates and in the level and slope components that the underlying hedging instrument alone may not span. Extending further, the three-swap case introduces in addition a 2y$\times$2y swap, which concentrates exposure at the front end of the curve, targeting short-term rate dynamics and improving control of near-dated shocks. 
Using short tenors and maturities that are far from these of the underlying for the additional hedging swaps allows
reducing correlation among hedging legs and protecting against a broader span of risks, 
yielding a more effective dynamic hedge. 

We use the following notation for rho-hedging. $\rho-X^{(i)}$ utilizes a single-swap hedge and neutralizes exposure to  the rate of factor $i$ at each rebalancing date. $\rho-(X^{(i)}, X^{(k)})$ hedges rely instead on two swaps and neutralize the exposure to the two factors $(i,k)$. Lastly, $\rho$ hedging neutralizes exposure to all three factors using three hedging swaps.

\begin{table}[ht]
\centering
\small
\setlength{\tabcolsep}{3pt}
\begin{tabular*}{\textwidth}{@{\extracolsep{\fill}}lcccccccc}
\toprule
Strategy & Mean & RMSE & RDR & CVaR$_{99\%}$ & P(HE $>$ 0) & HRR & TI & DTE \\
\midrule
\multicolumn{9}{c}{\textbf{Hedging with One Swap}} \\
\midrule
RL MSE         & -0.0033 & \textbf{0.0086} & 0.0042 & 0.0228 & 0.3220 & \textbf{0.8798} & 2.7623 & \textbf{0.0049} \\
RL DR          & -0.0052 & 0.0109 & \textbf{0.0032} & 0.0179 & 0.3092 & 0.8551 & 2.3316 & 0.0057 \\
RL CVaR        & \textbf{-0.0055} & 0.0128 & 0.0039 & \textbf{0.0157} & 0.3910 & 0.8251 & \textbf{2.2039} & 0.0066 \\
$\rho\!-\!X^{(1)}$ & -0.0046 & 0.0106 & 0.0038 & 0.0185 & 0.3380 & 0.8556 & 2.4602 & 0.0053 \\
$\rho\!-\!X^{(2)}$ & -0.0031 & 0.0147 & 0.0091 & 0.0394 & 0.3933 & 0.7838 & 2.4153 & 0.0109 \\
$\rho\!-\!X^{(3)}$ & -0.0038 & 0.0089 & 0.0039 & 0.0218 & \textbf{0.2965} & 0.8790 & 2.5955 & \textbf{0.0049} \\
\midrule
\multicolumn{9}{c}{\textbf{Hedging with Two Swaps}} \\
\midrule
RL MSE                      & -0.0025 & \textbf{0.0080} & 0.0042 & 0.0230 & 0.3413 & \textbf{0.8860} & 9.8791  & \textbf{0.0043} \\
RL DR                       & \textbf{-0.0087} & 0.0135 & \textbf{0.0022} & 0.0147 & \textbf{0.1729} & 0.8441 & 6.9082  & 0.0074 \\
RL CVaR                     & -0.0081 & 0.0151 & 0.0030 & \textbf{0.0131} & 0.3353 & 0.8088 & \textbf{4.4295}  & 0.0076 \\
$\rho\!-\!(X^{(1)},X^{(2)})$ & -0.0068 & 0.0120 & 0.0031 & 0.0175 & 0.2396 & 0.8508 & 7.6409  & 0.0063 \\
$\rho\!-\!(X^{(1)},X^{(3)})$ & -0.0037 & 0.0122 & 0.0059 & 0.0276 & 0.3933 & 0.8247 & 7.6791  & 0.0057 \\
$\rho\!-\!(X^{(2)},X^{(3)})$ & -0.0076 & 0.0132 & 0.0044 & 0.0238 & 0.2377 & 0.8380 & 6.6959  & 0.0083 \\
\midrule
\multicolumn{9}{c}{\textbf{Hedging with Three Swaps}} \\
\midrule
RL MSE         & -0.0014 & \textbf{0.0076} & 0.0045 & 0.0236 & 0.3981 & \textbf{0.8868} & 20.2907 & \textbf{0.0040} \\
RL DR          & \textbf{-0.0089} & 0.0137 & \textbf{0.0022} & 0.0145 & \textbf{0.1629} & 0.8438 & 10.6775 & 0.0076 \\
RL CVaR        & -0.0080 & 0.0143 & 0.0029 & \textbf{0.0131} & 0.3028 & 0.8209 & \textbf{4.1674}  & 0.0075 \\
$\rho$ Hedging & -0.0051 & 0.0106 & 0.0034 & 0.0182 & 0.2988 & 0.8603 & 6.7003  & 0.0051 \\
\bottomrule
\end{tabular*}
\caption{Performance metrics across reinforcement learning and rho-hedging strategies using one, two, and three swaps.}
\label{tab:rl_vs_rho_all_swaps_combined}
\end{table}

Table~\ref{tab:rl_vs_rho_all_swaps_combined} reports performance metrics for RL-based and rho-hedging strategies using one, two, and three swaps on an out-of-sample dataset. 

In the single-swap setting, RL methods consistently outperform traditional rho-hedging when the performance metric matches the objective function of the RL agent. The \textit{RL-MSE} strategy achieves the lowest RMSE (0.0086), highlighting its precision in tracking the target payoff. The \textit{RL-DR} specification provides the strongest downside protection, with the smallest root downside risk (0.0032). The \textit{RL-CVaR} approach offers the greatest reduction in tail exposure, recording the lowest CVaR$_{99\%}$ (0.0157). Furthermore, RL-CVaR strategies systematically have the lowest trading intensity, which would reduce trading costs in practice.
Among rho-hedging strategies, hedging the first factor provides the most risk protection, at least in terms of RDR and CVaR$_{99\%}$.

Adding a second swap substantially improves hedging performance for all strategies. RL strategies still retain a clear advantage. The \textit{RL-MSE} model again produces the highest risk reduction as measured by HRR (0.8860) and the lowest RMSE (0.0080), confirming better replication accuracy. 
The \textit{RL-DR} objective minimizes root downside risk (0.0022) and yields the lowest probability of underhedging (0.1729), while \textit{RL-CVaR} delivers superior tail performance with a CVaR$_{99\%}$ of 0.0131. 

Adding a third swap to the hedging portfolio only provides marginal additional benefits (if any) in terms of performance, both for RL methods and rho-hedging. For instance, the $\rho$-hedging strategy with three swaps underperforms the $\rho(X^{(1)},X^{(2)})$ in terms of DR and CVaR$_{99\%}$. This indicates that hedging effectiveness approaches a near-optimal level with two instruments. This can be explained by the high negative correlations between the level factor and the two others as seen in matrix $\rho$ from Appendix \ref{app:market_data}, which are respectively $-0.63$ and $-0.41$. While adding a new swap does not result in a significant improvement in performance, it notably increases trading intensity, especially for the RL-MSE and RL-DR strategies. There is thus insufficient de-correlation between factors to justify the use of a third hedging instrument. The inability of the third hedging swap to materially improve hedging performance is consistent with previous studies suggesting that increasing the number of hedging instruments beyond a certain point yields little incremental benefit and may raise instability \citep{fan2003hedging, driessen2003performance, fan2007pricing}.



\begin{figure}[ht]
  \centering
  \begin{tabular}{ccc}
    \begin{subfigure}[b]{0.32\textwidth}
      \includegraphics[width=\linewidth]{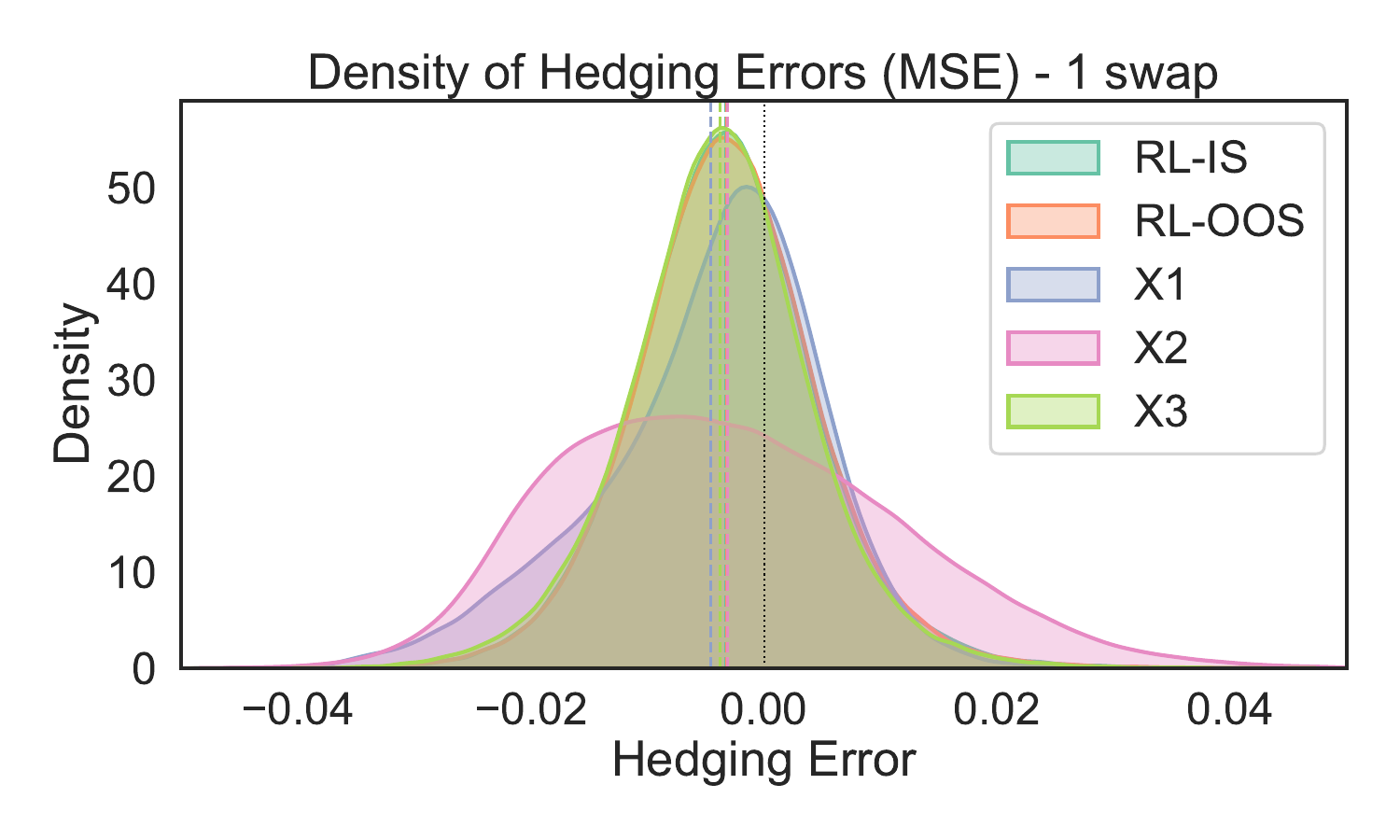}
    \end{subfigure} &
    \begin{subfigure}[b]{0.32\textwidth}
      \includegraphics[width=\linewidth]{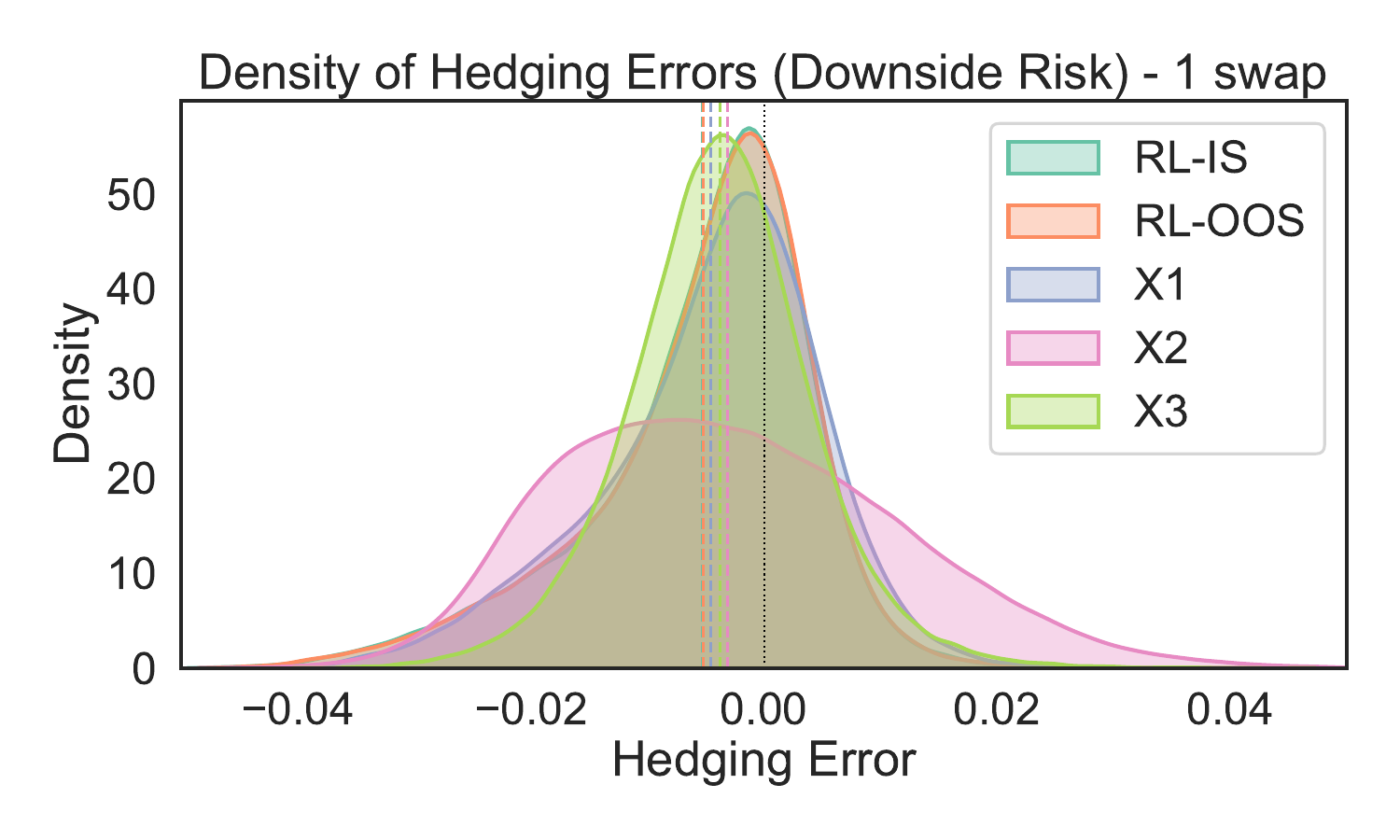}
    \end{subfigure} &
    \begin{subfigure}[b]{0.32\textwidth}
      \includegraphics[width=\linewidth]{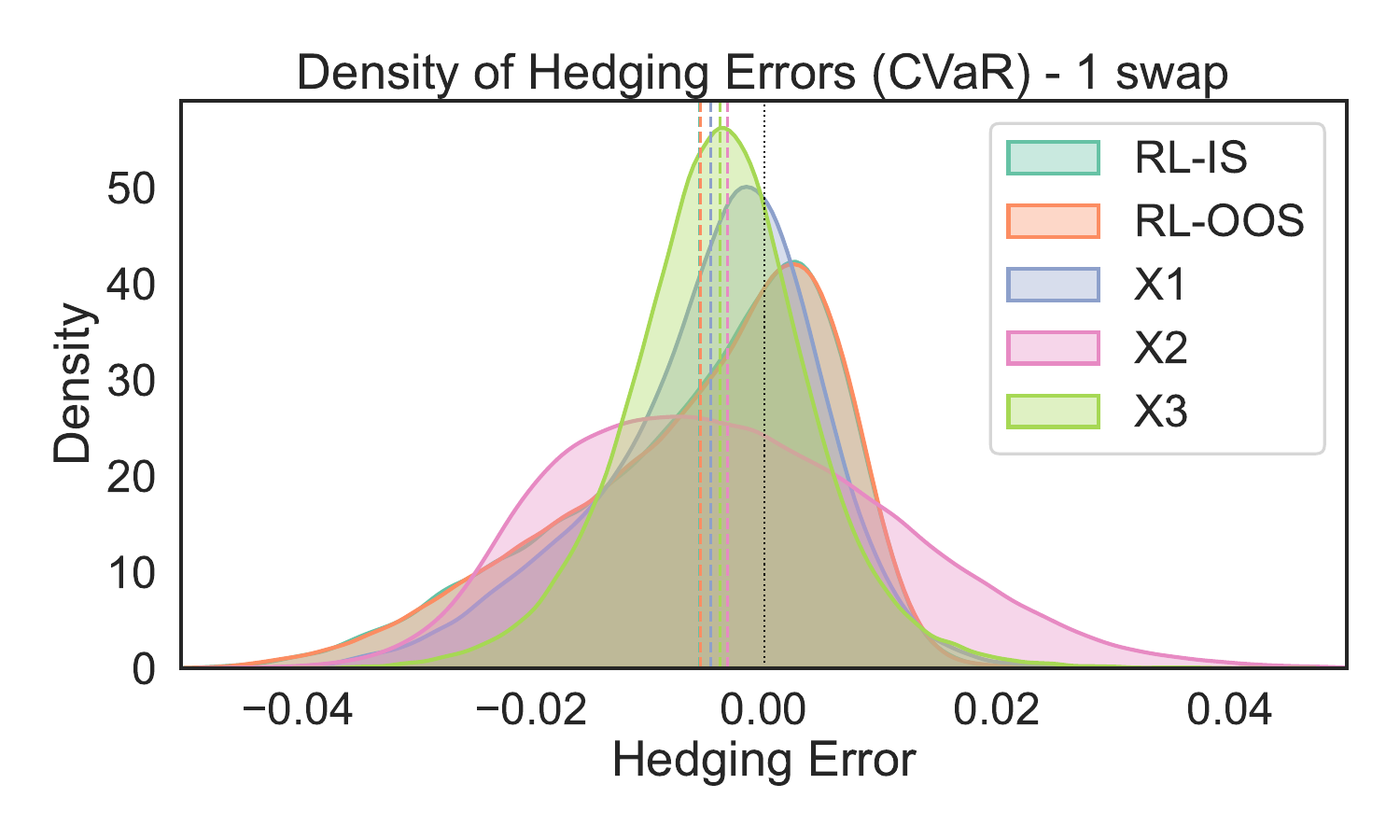}
    \end{subfigure} \\[1ex]

    \begin{subfigure}[b]{0.32\textwidth}
      \includegraphics[width=\linewidth]{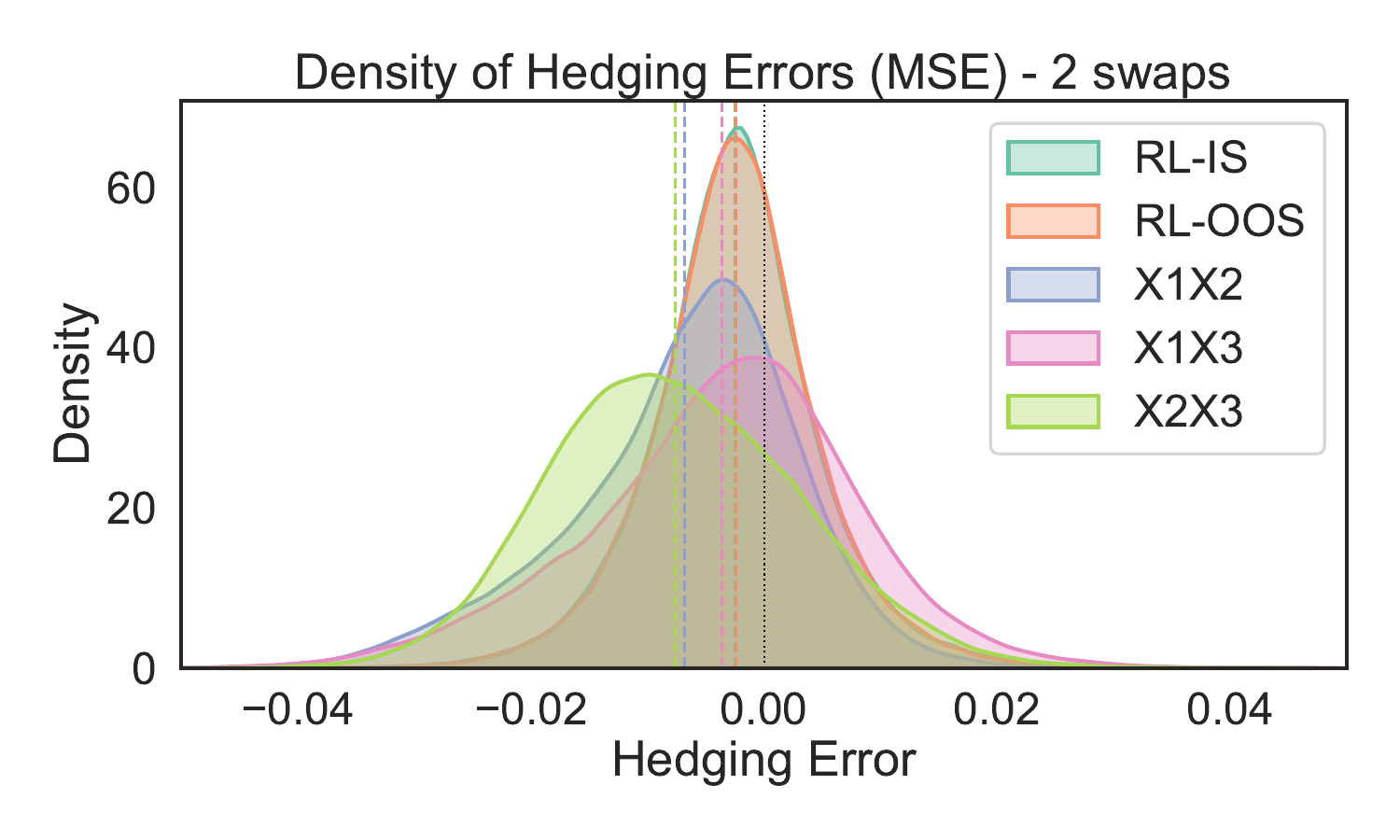}
    \end{subfigure} &
    \begin{subfigure}[b]{0.32\textwidth}
      \includegraphics[width=\linewidth]{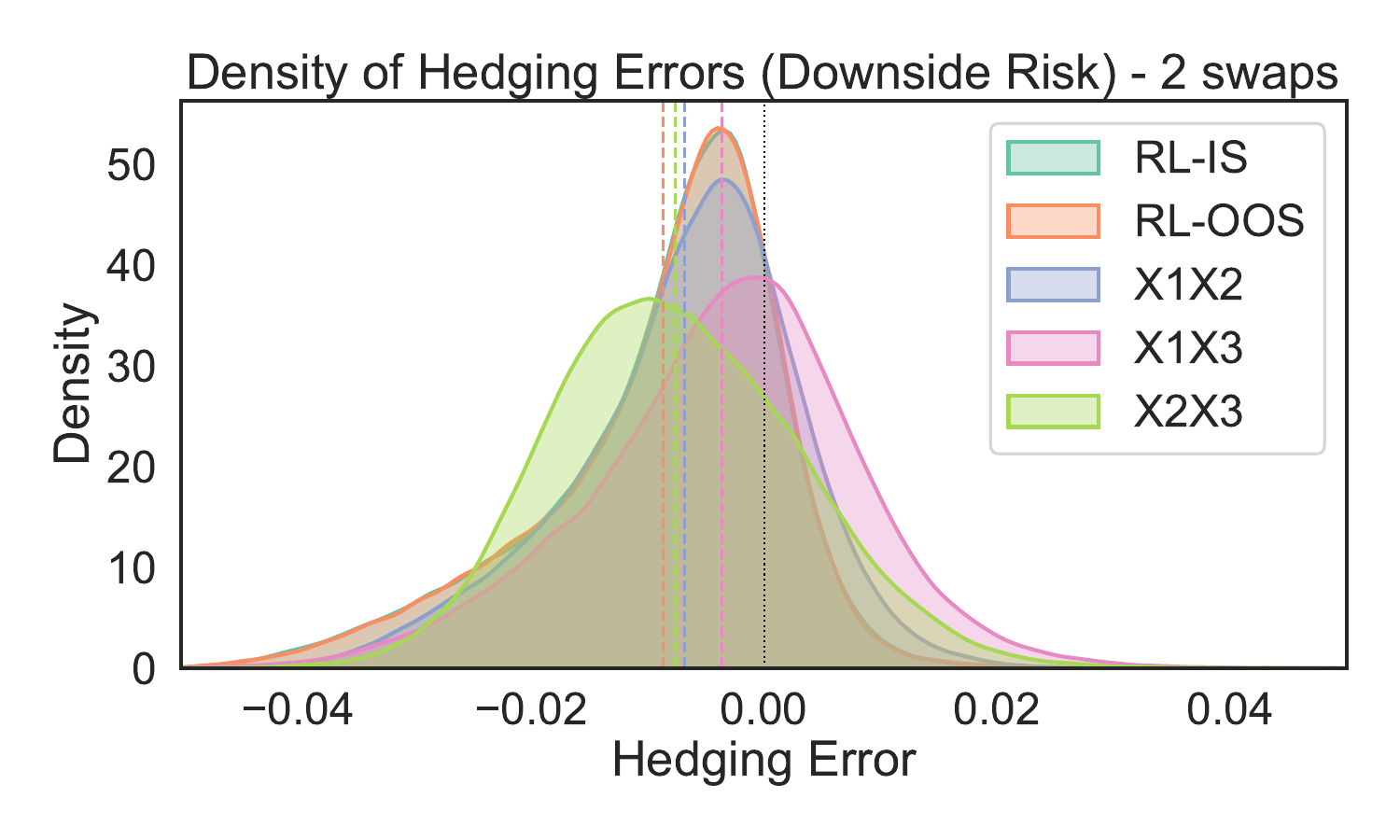}
    \end{subfigure} &
    \begin{subfigure}[b]{0.32\textwidth}
      \includegraphics[width=\linewidth]{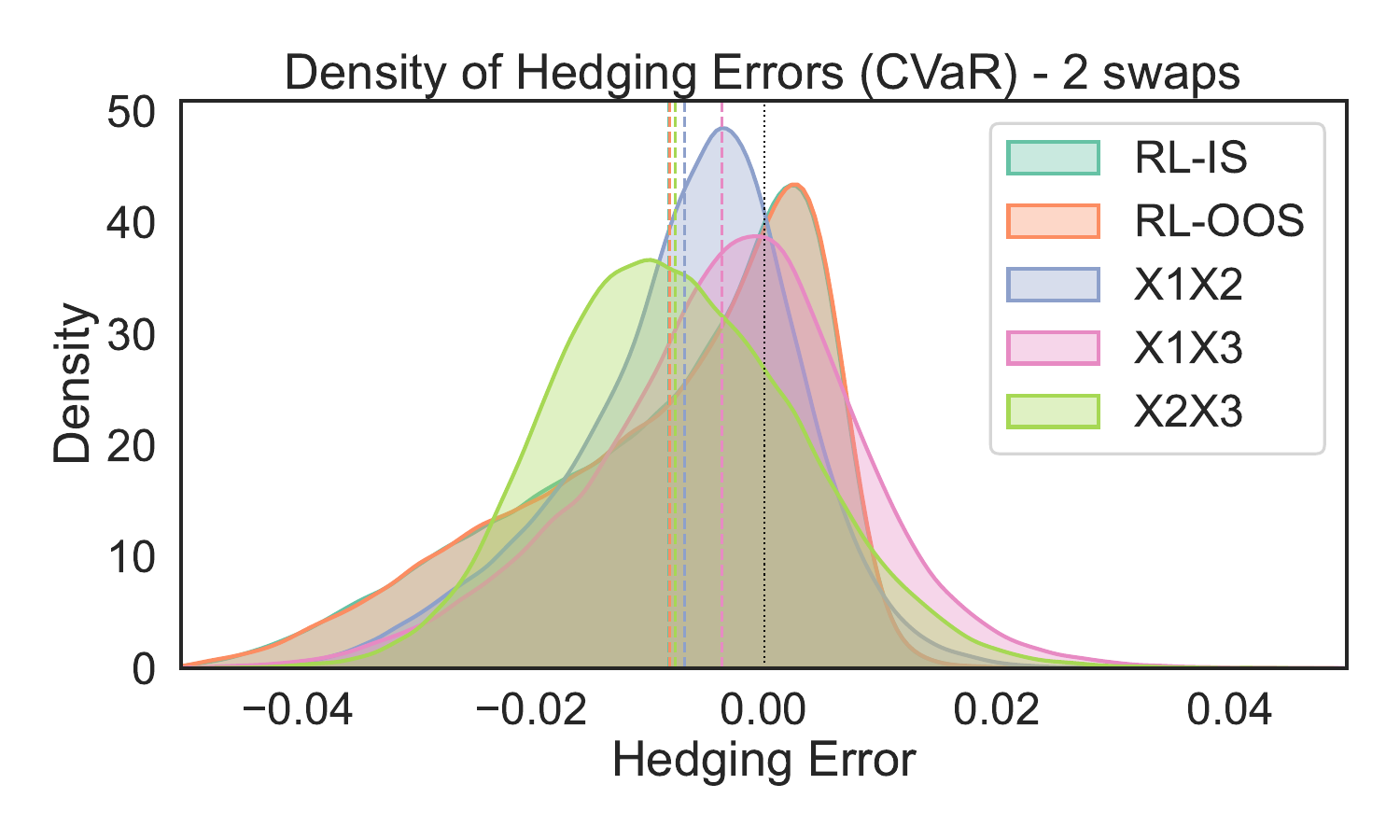}
    \end{subfigure} \\[1ex]

    \begin{subfigure}[b]{0.32\textwidth}
      \includegraphics[width=\linewidth]{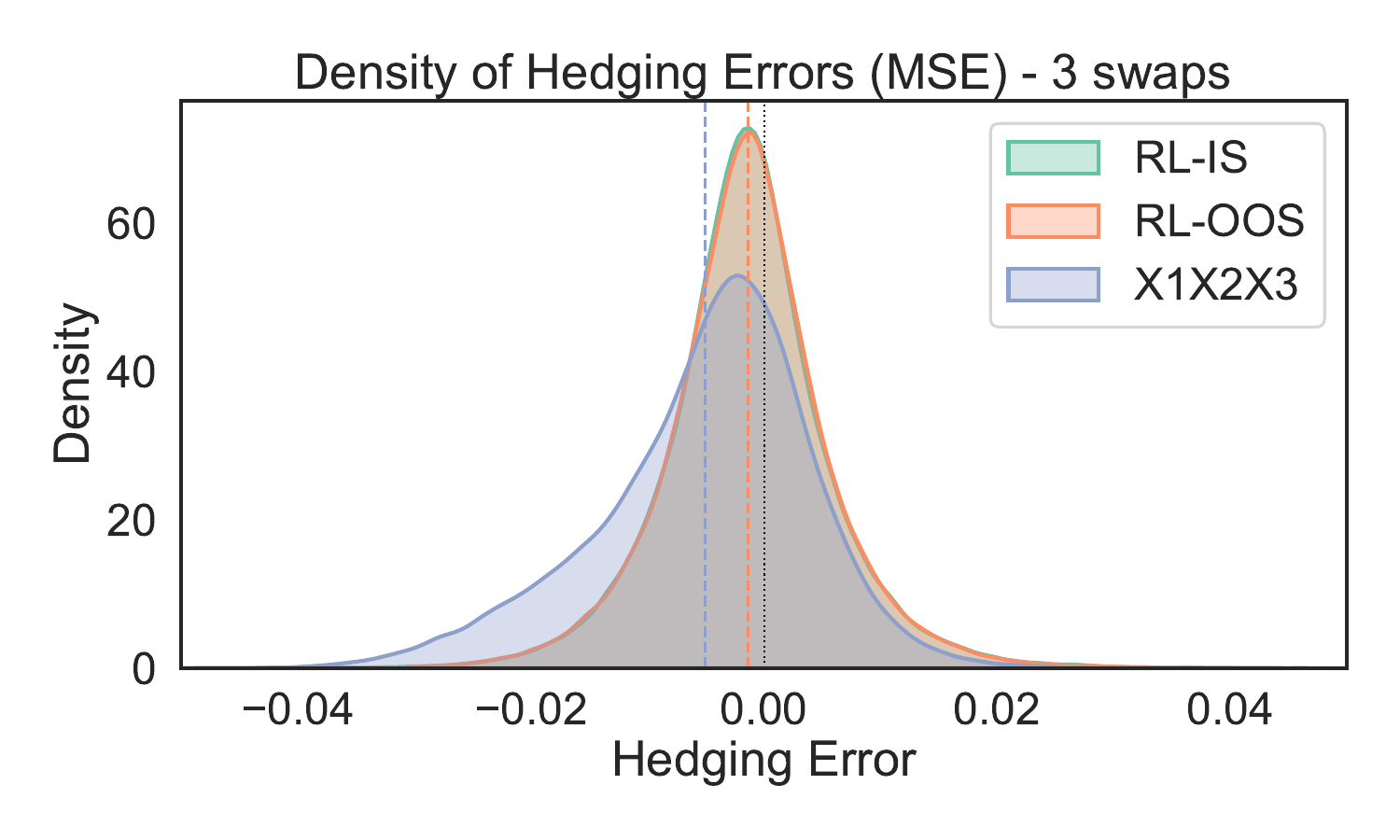}
      \caption{MSE}
    \end{subfigure} &
    \begin{subfigure}[b]{0.32\textwidth}
      \includegraphics[width=\linewidth]{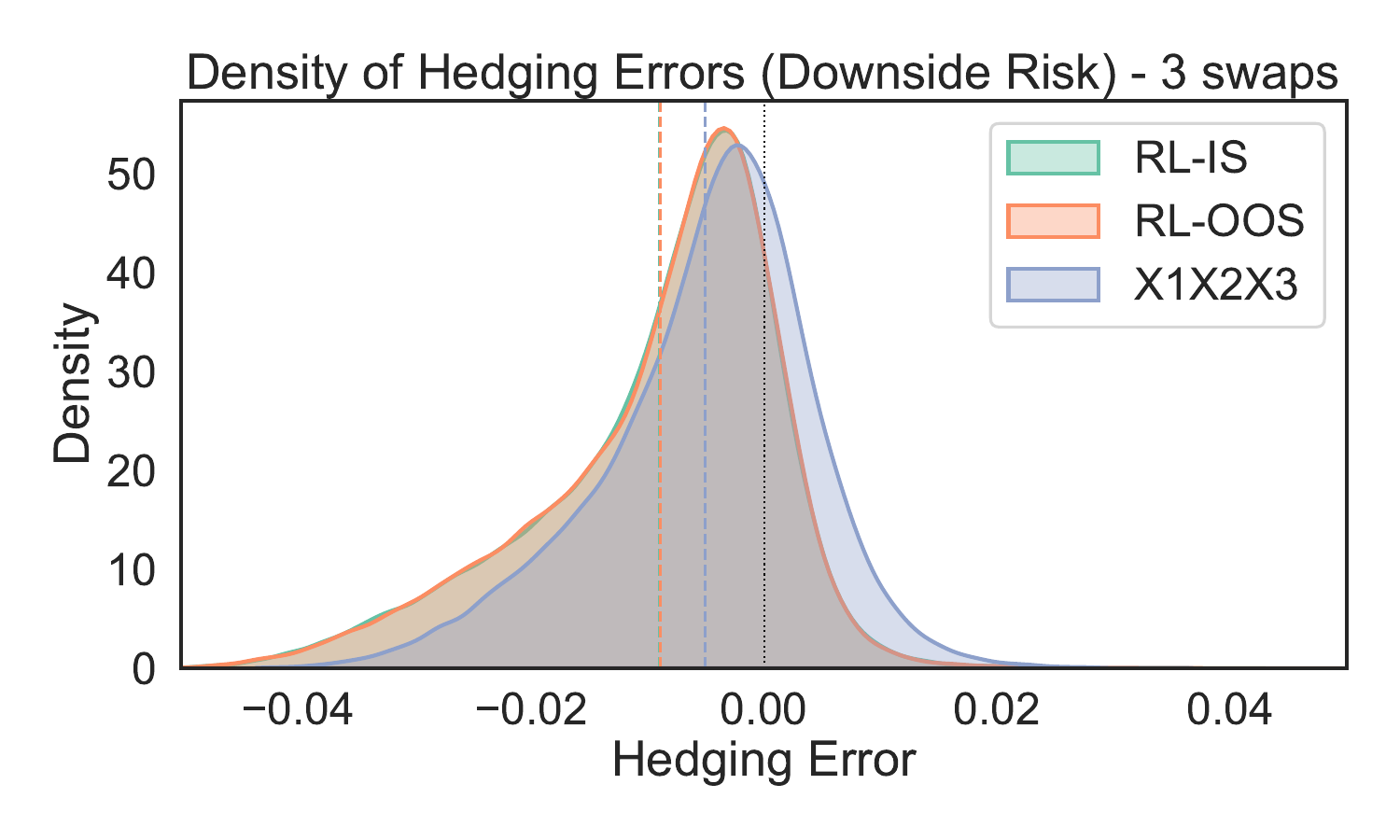}
      \caption{DR}
    \end{subfigure} &
    \begin{subfigure}[b]{0.32\textwidth}
      \includegraphics[width=\linewidth]{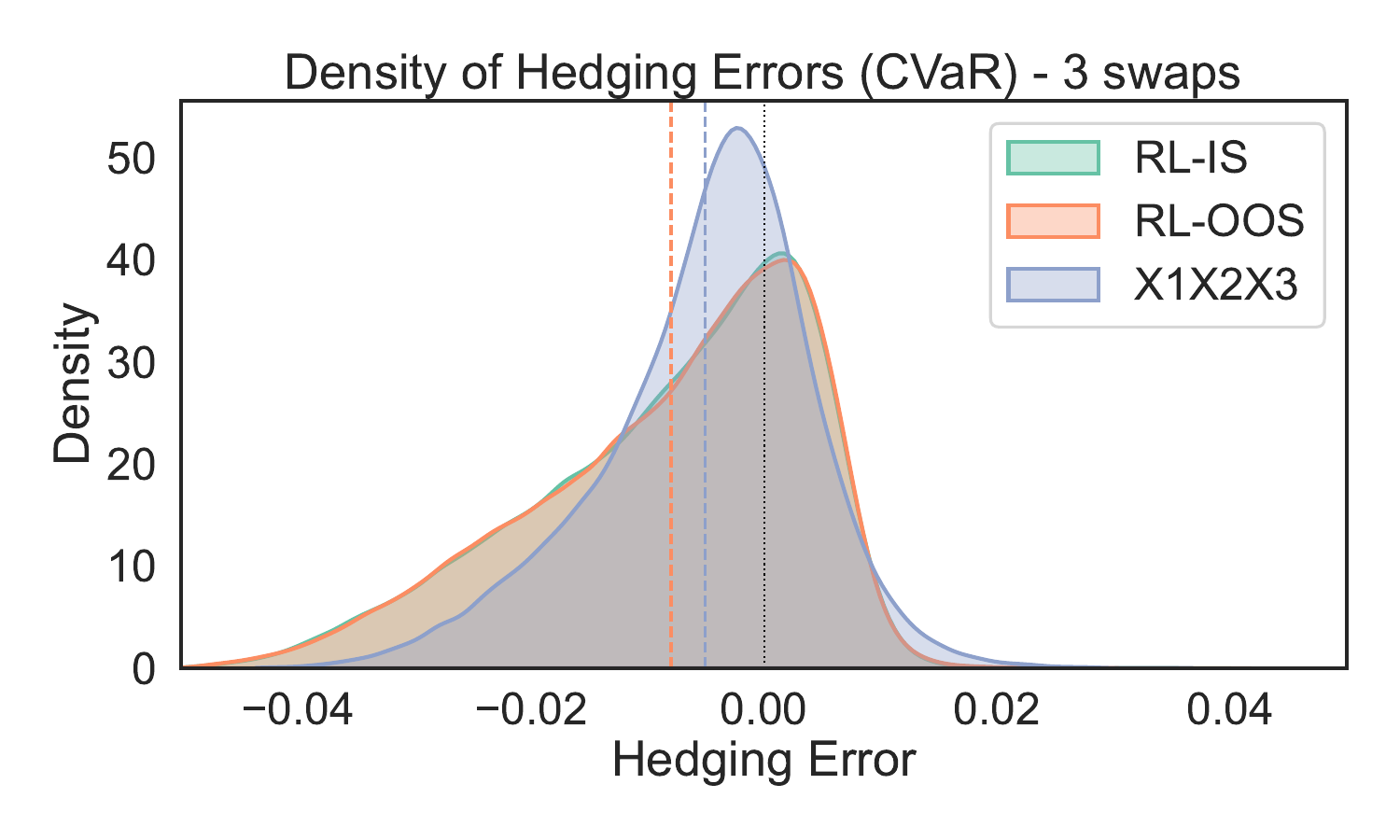}
      \caption{CVaR$_{99\%}$}
    \end{subfigure}
  \end{tabular}
  \caption{Comparison of kernel density estimates of in-sample (IS) and out-of-sample (OOS) hedging error distributions for RL-based strategies, and out-of-sample distributions for dynamic rho-hedge strategies, across different hedging portfolio compositions. The vertical lines indicate the mean of each distribution, while the black line indicates a null hedging error. The first row corresponds to hedging portfolios with one swap, the second row to two swaps, and the third row to three swaps.}
  \label{fig:hedging-errors-densities}
\end{figure}

Figure~\ref{fig:hedging-errors-densities} shows kernel density plots comparing the hedging error distributions of RL and rho-hedging strategies under the three objective risk measures for RL agents: MSE, DR, and CVaR$_{99\%}$. Strategies displayed use either one, two, or three swaps as hedging instruments. The RL strategies consistently produce tighter, well-centered error distributions, with very little difference between in-sample and out-of-sample results. This pattern indicates that the RL methods do not overfit the training data.
For the MSE objective especially, the RL error curves are symmetric, and they sharply peak near their mean,
indicating accurate and stable hedging. In contrast, the rho-hedging strategies often have wider distributions with heavier tails, and distribution modes are in several instances far from the mean.
This implies more frequent large errors and some bias in replication. Under the DR and CVaR objectives, RL methods exhibit thinner right tails, confirming that they manage risks related to extreme losses and downside fluctuations better. 

Overall, RL-based hedging consistently outperforms rho-hedging by delivering superior risk mitigation performance.

\subsection{Residual Sensitivity Analysis}

This section analyzes whether the outperformance of RL hedging strategies is partially due to
systematic exposure to some of the risk factors in order to harvest associated risk premia.
We conduct a residual exposure sensitivity analysis. Define, for any time step $t$, the residual vector, $\xi_t = [\xi_t^{(1)}, \xi_t^{(2)}, \xi_t^{(3)}]^\top$:
\[
\xi_t = A_t \phi_{t+1} - \nu_t, 
\]
where $\nu_t$ is the vector of the swaptions' sensitivities to the underlying risk factors,
and $A_t$ is a $3 \times M$ matrix of sensitivities of the available swaps in the replicating portfolio, where $M$ is the number of swaps in the replicating portfolio:
\begin{equation*}
    \Big(\nu_t \Big)_{i} = \frac{\partial \mathrm{PS_t}}{\partial X_t^{(i)}}, \quad \text{and} \quad \Big(A_t \Big)_{i,j} = \frac{\partial\mathrm{PFS_t^{(j)}}}{\partial X_t^{(i)}}, \quad \text{for} \quad i=1,2,3, \quad j=1, \dots, M.
\end{equation*}
By plotting the time series of the mean value across all paths of each component $\xi_t^{(k)}$, we can identify whether there is a systematic bias in any direction. If $\xi_t^{(k)}$ fluctuates around zero with low variance, the RL agent behaves similarly to a traditional rho-hedger.
Conversely, if $\xi_t^{(k)}$ consistently remains above (below) zero, it indicates that the RL agent systematically positions itself to benefit from an increase (decrease) 
in the value of factor $X^{(k)}$. This analysis helps to assess whether the observed outperformance of RL hedging strategies arises at least partially from directional exposure. 

We examine the agent’s behavior along two sets of paths: those that ultimately result in either (i) a positive payoff or (ii) a zero payoff. Focusing on these sets of scenarios separately provides insight into how the agent adjusts its actions based on the option moneyness as the portfolio approaches maturity. Paths associated with a high (low) payoff are associated with an increase (a decrease) in interest rates, specifically in the level factor. The value of payer swaps also increases when interest rates increase.

The three factors, corresponding respectively to the level ($X^{(1)}$), slope ($X^{(2)}$), and curvature ($X^{(3)}$) of the yield curve, are associated with distinct risk premia. The level (factor $X^{(1)}$) is associated with long-term interest rates and inflation expectations. This factor is linked to the term premium: investors typically require compensation for holding long-duration bonds. The slope (factor $X^{(2)}$) is associated with steepening/flattening of the yield curve and reflects the difference between short- and long-term rates. It is linked to carry/roll-down trade premium. The curvature (factor $X^{(3)}$) represents medium-term deviations, often tied to intermediate-term economic uncertainty or changing term premia. It is associated with convexity and volatility-of-volatility risk, where investors often demand compensation for bearing curvature exposure \citep{duffee2002term,diebold2006forecasting}.

In our model, we can measure risk premia by comparing the difference in the long-run averages of factors under the physical and risk-neutral measures, namely $\theta^\mathbb{P} - \theta^{\mathbb{Q}}$. For the slope and curvature factors, this difference points to premia of about \(-3.3\%\) and \(-2.6\%\), respectively. However, since the level factor process has a unit root under \(\mathbb{Q}\), the long-term mean is not $\theta^\mathbb{P}_1 - \theta^{\mathbb{Q}}_1$ and the latter quantity cannot be used to evaluate the risk premium.
Instead, the risk-neutral conditional expected value of $X^{(1)}_t$ given information at time zero stays fixed at \(X_0^{(1)}=-0.0312\). Conversely, the long-run mean for $X^{(1)}$ is $\theta^\mathbb{P}_1=0$ under the physical measure $\mathbb{P}$; we can thus conclude that there is a positive level risk premium at time $t=0$ given by $\theta^\mathbb{P}_1 - X_0^{(1)}=0.0312$.
Therefore, systematic long exposure to the slope or curvature factors entails effectively paying a risk premium, which is equivalent to buying insurance against unfavourable states. 
Cutting back on those exposures, or even holding negative exposure, in turn, reflects efforts to save on costs. On the other hand, holding long level exposure leads to collecting its positive premium. Conversely, reducing the exposure to the level factor reflects a desire to manage risk despite reducing prospective profits.



\begin{figure}[ht]
  \centering

  \includegraphics[width=0.9\textwidth]{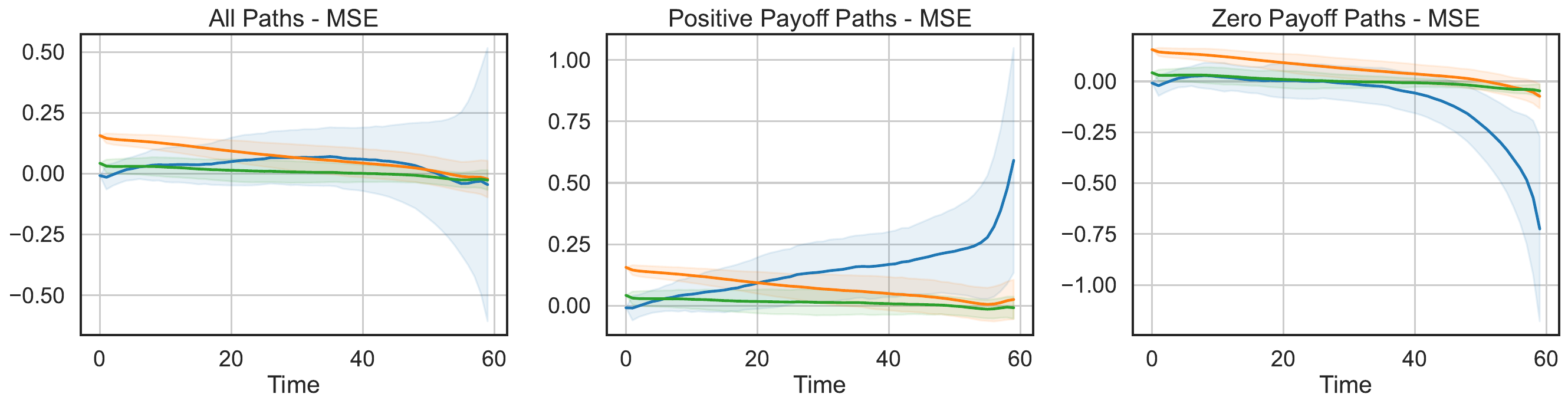}

  \includegraphics[width=0.9\textwidth]{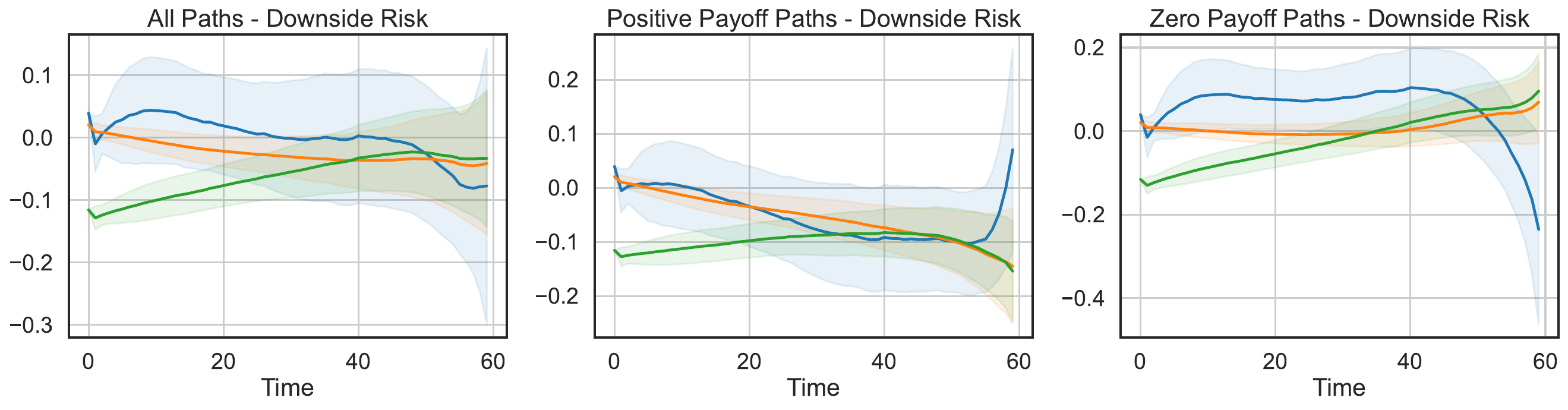}

  \includegraphics[width=0.9\textwidth]{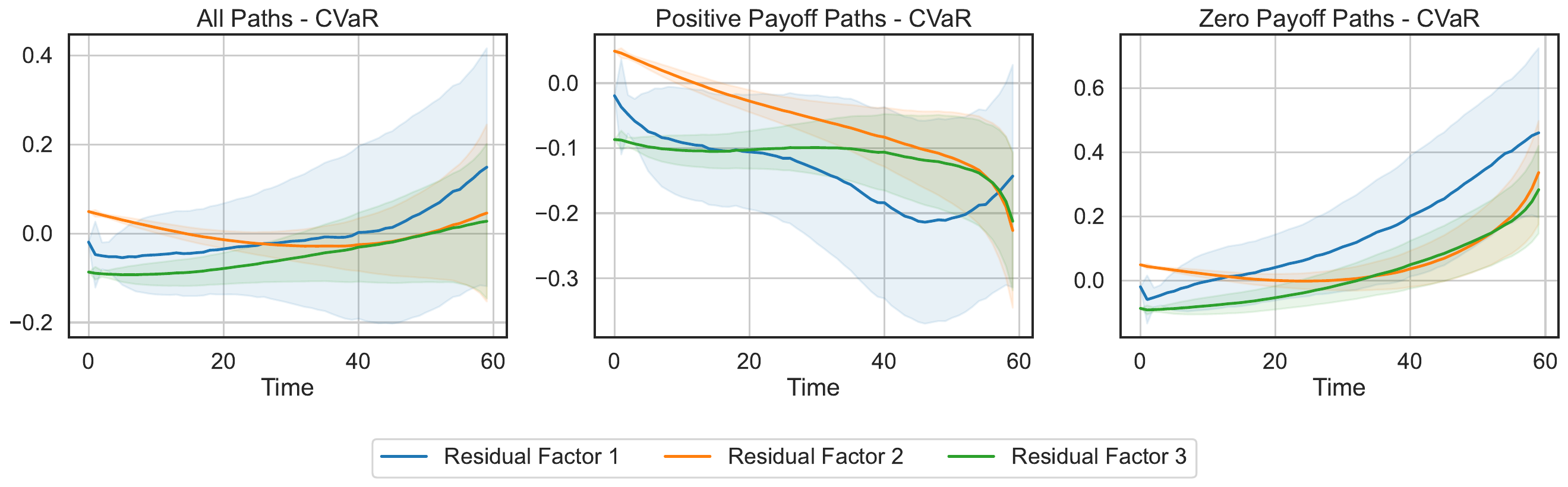}
  \caption{Residuals for 2-Swap Portfolio – Multiple Objectives. Shaded areas represent scaled standard deviation bands (0.5$\times$std) for visual clarity}
\label{fig:residuals_plots}
\end{figure}

In Figure~\ref{fig:residuals_plots}, we track the residual exposure to the three term structure factors when a short swaption position is hedged with two swaps. Results are presented for the RL agents using the MSE, DR, and CVaR$_{99\%}$ risk measures as objectives, and for rho-hedging. 

We observe that patterns in exposure to the slope and curvature are similar; for each RL objective function or scenario set considered, both tend  to increase or decrease simultaneously. This can be explained by these two factors having similary dynamics under our setting, with both being associated with negative risk premium, having a strong negative correlation with the level factor and being positively correlated together (0.2993). Contrarily, the level factor exhibits distictive dynamics; it is associated with a positive premium at $t=0$, it has a unit root under the risk-neutral measure and it is negatively correlated to the two other factors. This explains why, in several instances, exposure to the level factor displays different patterns than these associated with the curvature and slope factors.

The MSE agent responds symmetrically to hedging errors, aiming to minimize any deviation rather than guard only against downside or tail risk. When the payoff is likely to turn positive, it tends to hold a long position in the level factor, collecting the associated risk premium to help cover for such payoff and getting profits from hedging swaps. In contrast, when the trade is unlikely to finish in-the-money, as maturity approaches, the agent reduces its level exposure and tends to even go short, giving up potential upside in order to avoid hedging profits.
The average exposure to the other factors is modest; while the average curvature exposure is almost nil throughout the hedging horizon, the slope exposure moves towards zero as maturity approaches. 
Overall, the MSE agent strategy is simple and focused: it actively manages level risk to steer the hedging portfolio value and avoid non-zero hedging profits or shortfalls, while treating slope and curvature as secondary.

Risk exposures associated with the hedging strategy of the DR-RL agent depart from these of the MSE-RL agent; 
whereas the curvature exposure (green line) was almost null for the latter, large short (negative) exposures are observed for the former at earlier stages. This is reflective of asymmetric preferences of the RL agent who does not fear being penalized for large hedging profits potentially stemming from harvesting the (negative) risk premium on third factor through short exposure. Such harvested premium can help creating a buffer in early stages of the hedge to offset potential future losses. 

While CVaR-RL and DR-RL agents both have predominantly negative exposures within paths leading to a positive payoff, the exposure to the level factor in paths with a zero payoff is vastly different. For instance, the very large positive level and slope exposures observed in later stages for the CVaR-RL agent reflect the fact that, when the swaption is out-of-the-money (recall that the short rate is the sum of level and slope factors), the only way a tail-risk event can occur is through a very large spike in interest rates; positive exposure on the slope and level factors helps guarding against hikes in these two factors.

\subsection{Feature Importance}

To understand what drives RL hedging strategies, it is important to evaluate how much influence each model input has on the rebalacing decisions being applied. Analyzing feature importance allows us to see which of the state variables play a bigger role in hedging position adjustments, improving both transparency and interpretability. We assess the contribution of the term structure factors, namely the level, slope, and curvature of the yield curve, the hedging portfolio value, and the time-to-maturity, and examine how they shape the model’s hedging policy.

In our setting, we apply SHapley Additive exPlanations (SHAP) to the hedging portfolio positions generated by the RL policy. This framework follows the approach of \citet{vstrumbelj2014explaining}, later formalized by \citet{lundberg2017unified}. SHAP values decompose each hedging decision into contributions from the various state variables (i.e. features) by measuring each feature’s marginal contribution. To implement Shapley decompositions, we use the \texttt{shap} package in Python. Under this implementation, Shapley decomposition values are obtained without retraining the model on reduced sets of features. Instead, when computing predictions that rely on a subset of features, the model with the full set of features is applied with white noise values being used to substitute for values of features that are dropped. A sample of 3000 randomly chosen observations is used to report results.

\begin{figure}[ht]
    \centering

    \begin{subfigure}{0.45\textwidth}
        \centering
        \includegraphics[width=\linewidth]{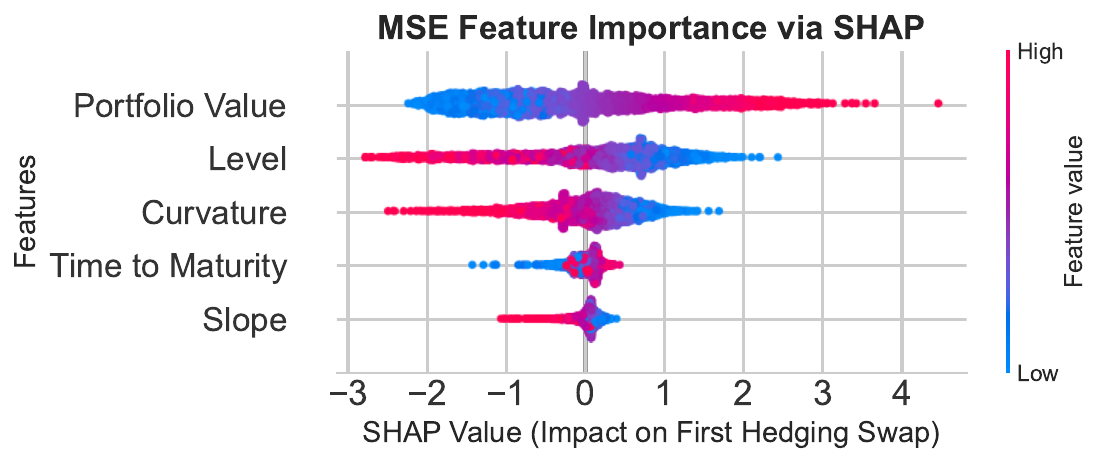}
    \end{subfigure}
    \begin{subfigure}{0.45\textwidth}
        \centering
        \includegraphics[width=\linewidth]{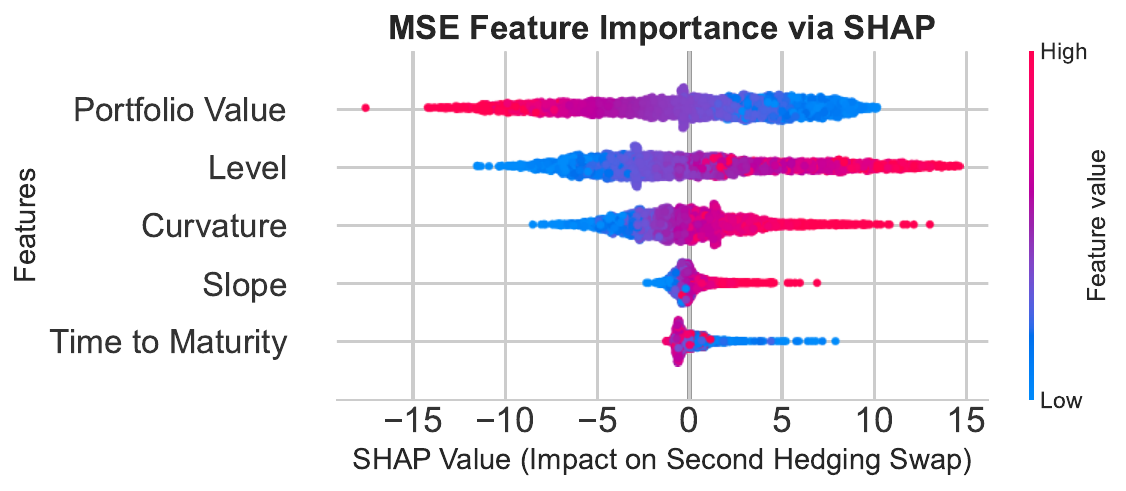}
    \end{subfigure}

    \vspace{0.3cm}
    \begin{subfigure}{0.45\textwidth}
        \centering
        \includegraphics[width=\linewidth]{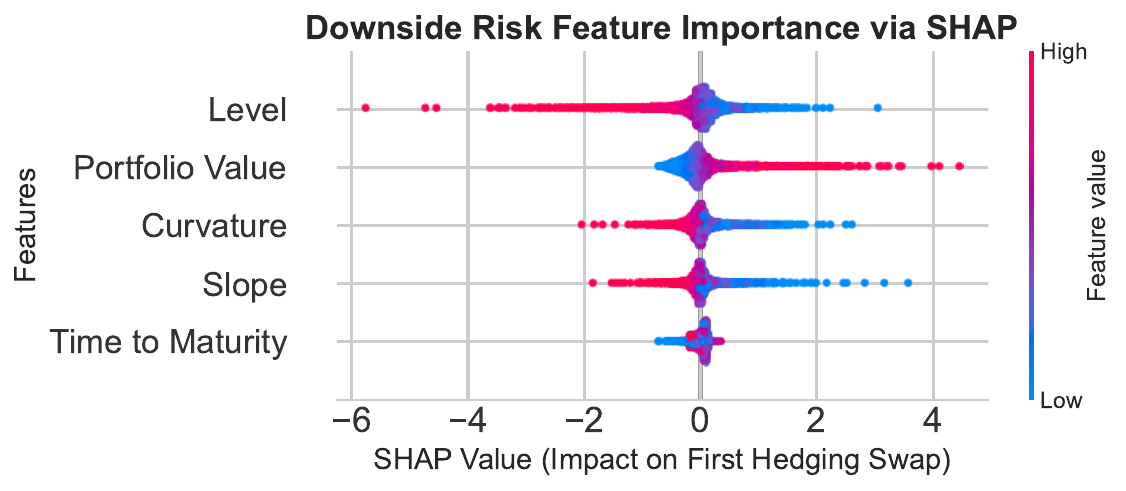}
    \end{subfigure}
    \begin{subfigure}{0.45\textwidth}
        \centering
        \includegraphics[width=\linewidth]{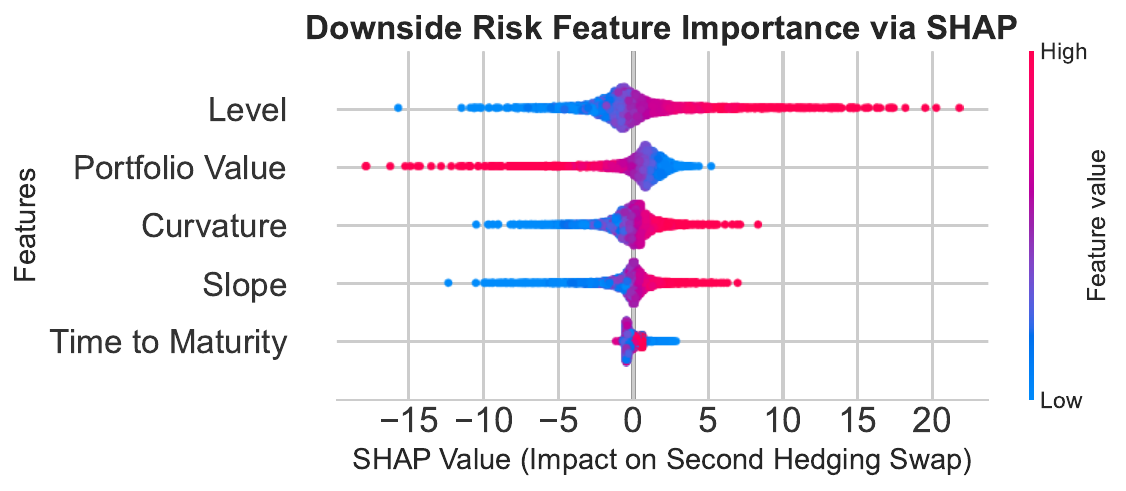}
    \end{subfigure}

    \vspace{0.3cm}
    \begin{subfigure}{0.45\textwidth}
        \centering
        \includegraphics[width=\linewidth]{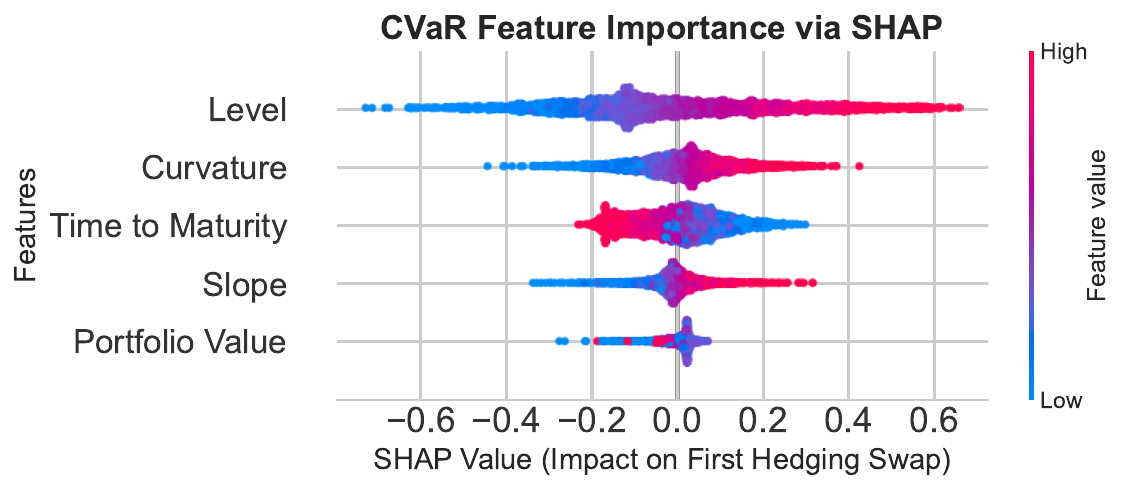}
    \end{subfigure}
    \begin{subfigure}{0.45\textwidth}
        \centering
        \includegraphics[width=\linewidth]{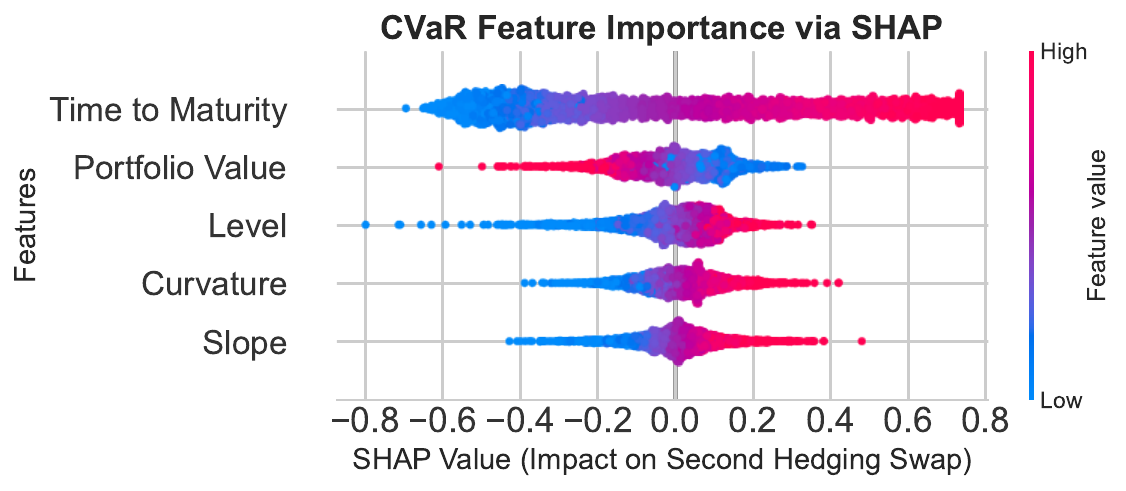}
    \end{subfigure}

    \caption{Shapley feature importance of hedge swap weights for two hedging strategies (MSE, DR, CVaR).}
    \label{fig:shap_beeswarm_grid}
\end{figure}

The resulting beeswarm plots shown in Figure~\ref{fig:shap_beeswarm_grid} provide an intuitive way to interpret these attributions. Please note that the x axis has a different scale for the various subpanels to enhance visibility. Each row represents one of the three hedging agents associated with the three considered objective functions. Results are displayed for the two-hedging-swap case, with left (right) columns depicting positions on the first (second) hedging swap.
The horizontal axis represents the SHAP value, i.e., the impact of a feature on the hedging position: positive values push the position upward, while negative values push it downward. Each dot corresponds to one observation. The color encodes the magnitude of the feature value, with red indicating high values and blue indicating low values; the combination of points color and of their horizontal axis value reveals the directional impact of the features. Features exhibiting a wider spread for SHAP values exert stronger influence on the RL model’s hedging decisions.
Taken together, these elements reveal not only which features matter most, but also how their realizations drive hedging position adjustments. 

Figure~\ref{fig:shap_beeswarm_grid} reveals several important insights. First, the SHAP value ranges differ across the three RL objectives, indicating that each agent adjusts its swap positions with a distinct level of sensitivity depending on the optimization criterion. Additionally, within each agent, the SHAP values for the first swap (the swap underlying the swaption) exhibit a much narrower range than those of the second swap.
This is because the tenor of the underlying swap is 10 years, whereas it is only 2 years for the second hedging swap; the price of the first swap exhibits much more variability and thus positions on the second swap must be on a higher scale to provide the same hedging effect.
Furthermore, the CVaR-RL agent displays a markedly narrower SHAP range compared with the MSE-RL and DR-RL agents, indicating far smaller adjustments in its hedge positions. This aligns with Table~\ref{tab:rl_vs_rho_all_swaps_combined}, where the CVaR-RL strategy exhibits the lowest trading intensity (TI). 
Comparing the MSE-RL and DR-RL agents, the SHAP distributions for MSE-RL are wider and more asymmetric, explaining why the MSE-RL strategy has substantially higher trading intensity than the DR-RL strategy. Nevertheless, for all variables and for both swaps, the overall direction of the impact of any given feature on hedging positions is the same for MSE-RL and DR-RL. Interestingly, a finding that is not reported numerically in this paper is that the DR-RL agent has on average a long position on the second payer hedging swap, whereas the average position on the second swap is negative (short) for the MSE-RL. Despite this gap, positions are impacted in the same direction by standalone changes in features.
Furtheremore, regardless of the agent and the feature, the impact on the first and on the second hedging swap positions are always in opposite directions, which means that the position on the second swap is used to partially offset some exposure generated by positions in the first swap. This reflects that the underlying hedging swap is the primary instrument replicating the swaption payoff. In contrast, the second swap serves mainly to absorb residual risks that the first swap cannot hedge, or to mitigate unwarranted risk exposure created by positions on the first swap. 



The level feature has very high variable importance for all three agents and both hedging swaps. The term structure level is the main driver of risk in our model. The portfolio value also is very important for the MSE-RL and DR-RL agents, whereas such variable is less impactful for the CVaR-RL agent. Interestingly, for the DR-RL agent, the impact of portfolio value is very asymmetric; when the hedger is ahead (i.e. large portfolio value) it can be aggressive in holding a larger underlying swap position to protect against potential hikes in interest rates. This is different for the MSE-RL agent who would be penalized for holding large positions on the first swap and collecting the level premium, since this would exacerbate large gains.
The inclusion of the portfolio value within state variables informing hedging decisions contributes to the out-performance of RL strategies over rho-hedging benchmarks, since the latter do not factor-in such information when making hedging decisions.
The time-to-maturity plays a more important role for the CVaR-RL agent; as maturity approaches the CVaR-RL hedger is better informed about whether the realized scenario is a tail risk event and must act more swiftly accordingly. The slope and curvature factors have more a modest impact that varies according to the agent's objective function.

\subsection{Market Sensitivity Analysis}
\label{subsec:market_sensitivity}

In this section, we examine the robustness of the hedging strategies to model misspecification. Specifically, we consider separately perturbations on two key sets of parameters of the yield curve factor process \eqref{eq:factor_p_measure}: the mean-reversion speed parameters $\kappa^{\mathbb{P}}$ and the long-run means $\theta^{\mathbb{P}}$. We then evaluate out-of-sample hedging performance of agents trained on original parameters when hedging on paths generated with perturbed parameters. Benchmarks also rely on hedging positions calculated with original parameters. 
We define perturbed parameter as
\[
\kappa' = c_\kappa \kappa^{\mathbb{P}}, \qquad \theta' = c_\theta \theta^{\mathbb{P}} ,
\]
with scaling constants set to $c_\kappa = c_\theta = 1.20$. 

Table~\ref{tab:kappa_sensitivity_rl_vs_rho_tc10bps} and Table~\ref{tab:theta_sensitivity_rl_vs_rho_tc10bps} report the performance of hedging strategies under perturbations respectively to the mean-reversion speed parameters $\kappa^{\mathbb{P}}$, and to the long-run mean parameters $\theta^{\mathbb{P}}$.

\begin{table}[ht]
\centering
\small
\setlength{\tabcolsep}{3pt}
\begin{tabular*}{\textwidth}{@{\extracolsep{\fill}}lcccccccc}
\toprule
Strategy & Mean & RMSE & RDR & CVaR$_{99\%}$ & P(HE $>$ 0) & HRR & TI & DTE \\
\midrule
\multicolumn{9}{c}{\textbf{Hedging with One Swap}} \\
\midrule
RL MSE         & -0.0031 & \textbf{0.0095} & 0.0053 & 0.0277 & 0.3392 & \textbf{0.8639} & 2.8249 & \textbf{0.0061} \\
RL DR          & -0.0052 & 0.0115 & \textbf{0.0041} & 0.0228 & 0.3126 & 0.8462 & 2.4137 & 0.0072 \\
RL CVaR        & \textbf{-0.0055} & 0.0128 & 0.0043 & \textbf{0.0191} & 0.3626 & 0.8257 & \textbf{2.2844} & 0.0080 \\
$\rho\!-\!X^{(1)}$ & -0.0045 & 0.0113 & 0.0048 & 0.0241 & 0.3309 & 0.8444 & 2.4602 & 0.0067 \\
$\rho\!-\!X^{(2)}$ & -0.0023 & 0.0156 & 0.0100 & 0.0404 & 0.4342 & 0.7683 & 2.4153 & 0.0117 \\
$\rho\!-\!X^{(3)}$ & -0.0037 & 0.0098 & 0.0050 & 0.0268 & \textbf{0.3114} & 0.8630 & 2.5955 & \textbf{0.0062} \\
\midrule
\multicolumn{9}{c}{\textbf{Hedging with Two Swaps}} \\
\midrule
RL MSE                      & -0.0024 & \textbf{0.0090} & 0.0054 & 0.0282 & 0.3528 & \textbf{0.8702} & 9.1072  & \textbf{0.0056} \\
RL DR                       & \textbf{-0.0091} & 0.0140 & \textbf{0.0025} & 0.0168 & \textbf{0.1740} & 0.8412 & 7.2776  & 0.0098 \\
RL CVaR                     & -0.0084 & 0.0152 & 0.0031 & \textbf{0.0156} & 0.3003 & 0.8110 & \textbf{4.5699}  & 0.0097 \\
$\rho\!-\!(X^{(1)},X^{(2)})$ & -0.0070 & 0.0127 & 0.0037 & 0.0205 & 0.2551 & 0.8414 & 7.6409  & 0.0086 \\
$\rho\!-\!(X^{(1)},X^{(3)})$ & -0.0036 & 0.0128 & 0.0067 & 0.0307 & 0.3826 & 0.8173 & 7.6791  & 0.0068 \\
$\rho\!-\!(X^{(2)},X^{(3)})$ & -0.0081 & 0.0137 & 0.0045 & 0.0260 & 0.2182 & 0.8349 & 6.6959  & 0.0102 \\
\midrule
\multicolumn{9}{c}{\textbf{Hedging with Three Swaps}} \\
\midrule
RL MSE         & -0.0012 & \textbf{0.0084} & 0.0056 & 0.0283 & 0.4084 & \textbf{0.8755} & 19.0079 & \textbf{0.0050} \\
RL DR          & \textbf{-0.0095} & 0.0141 & \textbf{0.0024} & 0.0163 & \textbf{0.1484} & 0.8442 & 11.2466 & 0.0101 \\
RL CVaR        & -0.0083 & 0.0144 & 0.0031 & \textbf{0.0156} & 0.2782 & 0.8238 & \textbf{4.2854}  & 0.0096 \\
$\rho$ Hedging & -0.0051 & 0.0113 & 0.0044 & 0.0227 & 0.3071 & 0.8494 & 6.7003  & 0.0069 \\
\bottomrule
\end{tabular*}
\caption{Performance metrics under \(\kappa^{\mathbb{P}}\)-shocks for RL and \(\rho\)-hedging strategies using one, two, and three swaps. }
\label{tab:kappa_sensitivity_rl_vs_rho_tc10bps}
\end{table}

The impact of parameter misspecification differs between the two perturbation scenarios. When $\kappa^{\mathbb{P}}$ is perturbed, the deterioration in hedging performance is more important for rho-hedging strategies than for deep hedges. For instance, in the case of hedging with three swaps, the absolute difference between the RDR of the RL-DR trading strategy with shocks on $\kappa$ (displayed in Table~\ref{tab:kappa_sensitivity_rl_vs_rho_tc10bps}) and that without such shocks (see Table \ref{tab:rl_vs_rho_all_swaps_combined}) is only $0.0002$. Similarly, the absolute change in $CVaR_{99\%}$ for the RL-CVaR agent is $0.0025$. Conversely the rho-hedges show higher sensitivity to parameter perturbation, with corresponding RDR and $CVaR_{99\%}$ differences being $0.0010$ and $0.0045$, respectively. 
Errors in the speed of mean-reversion mainly affect the persistence of market shocks, which RL policies are better able to absorb.

Perturbations to $\theta^{\mathbb{P}}$ lead to more pronounced degradation of performance across all metrics. Shifting the long-run mean introduces a structural drift in the factor dynamics, which results in larger MSE, downside risk and $\mathrm{CVaR}_{99\%}$. Both RL and $\rho$-hedges deteriorate under these conditions, but  the RL agents still have higher performance than the traditional first-order sensitivity-based strategies, e.g. $0.0057$ versus $0.0070$ for RDR and $0.0282$ versus $0.0294$ for $\mathrm{CVaR}_{99\%}$ in the three-hedging-swaps case.

RL strategies consistently deliver lower tail risk, and better hedging risk reduction than rho-hedges under perturbed parameters. 
The main reason why the hedging models are less sensitive to changes in $\kappa^{\mathbb{P}}$ is that altering the speed of mean reversion does not drastically modify the average shape of the factor paths; they still revert toward the same long‐run mean. In contrast, changing the long‐run mean $\theta^{\mathbb{P}}$ shifts the entire yield curve factor long-term levels; the effect of altered risk factor premia to which hedging portfolios are exposed compounds over time, which then leads to a substantial drop in performance.

\begin{table}[ht]
\centering
\small
\setlength{\tabcolsep}{3pt}
\begin{tabular*}{\textwidth}{@{\extracolsep{\fill}}lcccccccc}
\toprule
Strategy & Mean & RMSE & RDR & CVaR$_{99\%}$ & P(HE $>$ 0) & HRR & TI & DTE \\
\midrule
\multicolumn{9}{c}{\textbf{Hedging with One Swap}} \\
\midrule
RL MSE         & -0.0028 & 0.0134 & 0.0080 & 0.0352 & 0.4346 & 0.8021 & 2.7291 & \textbf{0.0114} \\
RL DR          & -0.0044 & 0.0148 & \textbf{0.0072} & 0.0313 & 0.4347 & 0.7864 & 2.3318 & 0.0124 \\
RL CVaR        & \textbf{-0.0047} & 0.0153 & 0.0071 & \textbf{0.0263} & \textbf{0.4135} & 0.7800 & \textbf{2.2091} & 0.0127 \\
$\rho\!-\!X^{(1)}$ & -0.0039 & 0.0142 & 0.0073 & 0.0295 & 0.4267 & 0.7939 & 2.4602 & 0.0120 \\
$\rho\!-\!X^{(2)}$ & -0.0007 & 0.0187 & 0.0129 & 0.0445 & 0.5066 & 0.7181 & 2.4153 & 0.0132 \\
$\rho\!-\!X^{(3)}$ & -0.0032 & \textbf{0.0133} & 0.0073 & 0.0321 & 0.4299 & \textbf{0.8064} & 2.5955 & 0.0117 \\
\midrule
\multicolumn{9}{c}{\textbf{Hedging with Two Swaps}} \\
\midrule
RL MSE                      & -0.0021 & \textbf{0.0134} & 0.0084 & 0.0362 & 0.4602 & \textbf{0.8216} & 9.5996  & \textbf{0.0110} \\
RL DR                       & \textbf{-0.0073} & 0.0162 & \textbf{0.0059} & 0.0286 & 0.3713 & 0.8055 & 6.9664  & 0.0145 \\
RL CVaR                     & -0.0068 & 0.0169 & 0.0064 & \textbf{0.0246} & 0.3720 & 0.7922 & \textbf{4.4346}  & 0.0141 \\
$\rho\!-\!(X^{(1)},X^{(2)})$ & -0.0055 & 0.0151 & 0.0068 & 0.0290 & 0.3931 & 0.8109 & 7.6409  & 0.0133 \\
$\rho\!-\!(X^{(1)},X^{(3)})$ & -0.0028 & 0.0154 & 0.0092 & 0.0359 & 0.4353 & 0.7960 & 7.6791  & 0.0117 \\
$\rho\!-\!(X^{(2)},X^{(3)})$ & -0.0072 & 0.0155 & 0.0066 & 0.0344 & \textbf{0.2756} & 0.8157 & 6.6959  & 0.0151 \\
\midrule
\multicolumn{9}{c}{\textbf{Hedging with Three Swaps}} \\
\midrule
RL MSE         & -0.0016 & \textbf{0.0130} & 0.0083 & 0.0351 & 0.4791 & \textbf{0.8273} & 19.9834 & \textbf{0.0107} \\
RL DR          & \textbf{-0.0075} & 0.0160 & \textbf{0.0057} & 0.0282 & \textbf{0.3544} & 0.8101 & 10.7588 & 0.0147 \\
RL CVaR        & -0.0067 & 0.0162 & 0.0062 & \textbf{0.0246} & 0.3667 & 0.8026 & \textbf{4.1602}  & 0.0140 \\
$\rho$ Hedging & -0.0045 & 0.0144 & 0.0070 & 0.0294 & 0.4141 & 0.8169 & 6.7003  & 0.0124 \\
\bottomrule
\end{tabular*}
\caption{Performance metrics under \(\theta^{\mathbb{P}}\)-shocks for RL and \(\rho\)-hedging strategies using one, two, and three swaps. }
\label{tab:theta_sensitivity_rl_vs_rho_tc10bps}
\end{table}

\section{Conclusion \label{sec:conclusion}}

In this paper, we set out to investigate whether RL can provide a more effective and robust framework to produce dynamic hedges of swaption positions compared to traditional sensitivity-based  rho-hedging.
The analysis yields several key insights. First, based on our three-factor arbitrage-free dynamic Nelson-Siegel model setting, we find that the incremental benefit of adding a second swap into the hedging portfolio is substantial, while further improvements from the inclusion of a third instrument are modest. While the underlying asset swap is the primary instrument that is relied upon for replication, the second swap allows adjusting residual exposures, e.g. mitigating unwarranted exposure created by positions on the first swap.
Our approach incorporates multiple possible objective functions, namely mean squared error, downside risk, and CVaR. We highlight that different risk preferences translate into distinct hedging styles, with dynamically-adjusted risk exposure to the various term structure factors exhibiting different patterns across the various RL objective functions. A variable importance assessment highlights that the level risk factor is the most impactful out of the three term structure factors in terms of hedging decisions. The hedging portfolio value, which is not used by benchmark strategies, also influences hedging decisions considerably, at least for RL agents using MSE or downside risk as their objective. Model misspecification tests, in which hedging performance is assessed on interest rate paths generated with perturbed parameters, highlight the resilience of the RL strategies which still out-perform benchmarks. Taken together, these findings suggest that reinforcement learning offers a compelling framework for the dynamic hedging of swaptions.

\bibliographystyle{apalike}
\bibliography{references}  


\appendix
\section{Zero-Coupon price} \label{app:zero_coupon_price}
As shown in \cite{eghbalzadeh2024discrete}, the time-$t$ price of a risk-free zero-coupon bond paying one dollar on maturity $T > t$ is, under the such model,
\begin{equation*}
P(t,T) = A_{\tau} \exp\left[-\Delta B_{\tau}^\top X_t\right], 
\end{equation*}
where $\tau = T - t$, $B_{\tau} = \left[ B^{(1)}_{\tau},  B^{(2)}_{\tau},  B^{(3)}_{\tau} \right]^\top$ and
\begin{equation*}
B^{(1)}_{\tau} = \tau, \quad 
B^{(2)}_{\tau} = \frac{1 - (1 - \lambda)^{\tau}}{\lambda}, \quad 
B^{(3)}_{\tau} = \frac{1 - (1 - \lambda)^{\tau - 1}}{\lambda} - (\tau - 1)(1 - \lambda)^{\tau - 1},
\end{equation*}

\begin{equation*}
    \log A_\tau = -\Delta \theta_2^{\mathbb{Q}} (B_\tau^{(1)} - B_\tau^{(2)}) 
+ \Delta \theta_3^{\mathbb{Q}} B_\tau^{(3)} 
+ \frac{1}{2} \Delta^2 v_\tau,
\end{equation*}
with

\lefteq{
v_\tau = \left( \sum_{i=1}^3 \sum_{j=1}^3 v_\tau^{(i,j)} \right),
}

\lefteq{
v_\tau^{(1,1)} = \Sigma_{1,1}^2 \frac{\tau(\tau - 1)(2\tau - 1)}{6},
}
\lefteq{
v_\tau^{(2,2)} = \frac{\Sigma_{2,2}^2}{\lambda^2} \left( \tau - 2 \left[ \frac{1 - (1 - \lambda)^\tau}{\lambda} \right] 
+ \frac{1 - (1 - \lambda)^{2\tau}}{1 - (1 - \lambda)^2} \right),
}
\lefteq{
v_\tau^{(3,3)} = \mathds{1}_{\{\tau > 1\}} \frac{\Sigma_{3,3}^2}{\lambda^2} \left[ 
\tau - 2 + \zeta_0((1 - \lambda)^2, \tau - 1) + \lambda^2 \zeta_2((1 - \lambda)^2, \tau - 1) \right.
}
\[
\left. - 2\zeta_0((1 - \lambda), \tau - 1) - 2\lambda \zeta_1((1 - \lambda), \tau - 1) 
+ 2\lambda \zeta_1((1 - \lambda)^2, \tau - 1) \right],
\]

\lefteq{
v_\tau^{(1,2)} = v_\tau^{(2,1)} = \rho_{1,2} \Sigma_{1,1} \Sigma_{2,2} \frac{1}{\lambda} 
\left( \frac{\tau(\tau - 1)}{2} - \zeta_1((1 - \lambda), \tau) \right),
}

\lefteq{
v_\tau^{(1,3)} = v_\tau^{(3,1)} = \mathds{1}_{\{\tau > 1\}} \rho_{1,3} \Sigma_{1,1} \Sigma_{3,3} \frac{1}{\lambda}
\Big[
\frac{\tau(\tau - 1)}{2} - 1 - \zeta_0((1 - \lambda), \tau - 1) }
\[ - (\lambda + 1) \zeta_1((1 - \lambda), \tau - 1)
- \lambda \zeta_2((1 - \lambda), \tau - 1)
\Big],
\]

\lefteq{
v_\tau^{(2,3)} = v_\tau^{(3,2)} = \mathds{1}_{\{\tau > 1\}} \rho_{2,3} \Sigma_{2,2} \Sigma_{3,3}
\left(
\frac{
\tau - 2 - (2 - \lambda) \zeta_0((1 - \lambda), \tau - 1) + (1 - \lambda) \zeta_0((1 - \lambda)^2, \tau - 1)
}{\lambda^2}
\right.
}
\[
\left.
+ \frac{
- \zeta_1((1 - \lambda), \tau - 1) + (1 - \lambda) \zeta_1((1 - \lambda)^2, \tau - 1)
}{\lambda}
\right),
\]

\lefteq{
\zeta_0(r, \tau) \equiv \sum_{u=1}^{\tau - 1} r^u = \frac{r - r^{\tau}}{1 - r},
}

\lefteq{
\zeta_1(r, \tau) \equiv \sum_{u=1}^{\tau - 1} u r^u = \frac{r - \tau r^{\tau} + (\tau - 1) r^{\tau + 1}}{(1 - r)^2}, 
}

\lefteq{
\zeta_2(r, \tau) \equiv \sum_{u=1}^{\tau - 1} u^2 r^u 
= \frac{-(\tau - 1)^2 r^{\tau + 2} + (2\tau^2 - 2\tau - 1) r^{\tau + 1} - \tau^2 r^{\tau} + r^2 + r}{(1 - r)^3}. 
}

\section{Factor Dynamics under the $T$-forward Measure} \label{app:forward_measure}

Results from this appendix are drawn from \cite{godin2023pricing}. The $T$-forward measure, denoted by $\mathbb{Q}^T$, is defined using the price of a zero-coupon bond maturing at time $T$ as the numéraire. The Radon–Nikodym derivative for changing from the risk-neutral measure $\mathbb{Q}$ to $\mathbb{Q}^T$ is given by
\[
\frac{d\mathbb{Q}^T}{d\mathbb{Q}} = \frac{D(0, T)}{P(0, T)}.
\]
Let $\tau = T - t$. Under the $T$-forward measure, the innovation process satisfies
\[
Z_{t+1}^T = Z_{t+1}^{\mathbb{Q}} + \Delta \Sigma B_{\tau - 1},
\]
and the conditional distribution of $Z_{t+1}^T$ given $\mathcal{F}_t$ is Gaussian with zero mean and covariance matrix $\rho$. As shown in Proposition 2.1 and Corollary 2.1 of \cite{godin2023pricing}, the sequence $\{Z_j^T\}_{j=1}^T$ is conditionally independent under $\mathbb{Q}^T$.

Using this, the dynamics of the term structure factors under the $T$-forward measure can be written in the affine form
\begin{equation*}
\label{eq:factor_forward}
X_{t+1} = X_t - \eta_t^T + \kappa^T(\theta^T - X_t) + \Sigma Z_{t+1}^T,
\end{equation*}
where the parameters are defined as
\[
\theta^T = \theta^{\mathbb{Q}}, \quad \kappa^T = \kappa^{\mathbb{Q}}, \quad \eta_t^T = \Delta \Sigma \rho \Sigma B_{\tau - 1}.
\]

Finally, Proposition 2.2 in the same paper states that for any $t + n \leq T$, the future factor values $X_{t+n}$, conditionally on $\mathcal{F}_t$, follow a multivariate Gaussian distribution under $\mathbb{Q}^T$ with explicitly computable mean vector and covariance matrix whose formulas are given in \cite{godin2023pricing}. This tractability plays a crucial role in pricing and hedging derivatives such as swaptions under the DTAFNS model.

\section{Model Parameter Setup} \label{app:market_data}

We use the calibrated DTAFNS model parameters proposed in \cite{eghbalzadeh2024discrete}. They perform a joint estimation of parameters from measures $\mathbb{P}$ and $\mathbb{Q}$, which gives the following estimates:
\begin{itemize}
    \item \textbf{Shape parameter:} $\lambda = 0.0233$.
\end{itemize}


\begin{itemize}
    \item \textbf{Long-term means:}
    \[
    \theta^{\mathbb{P}} =
    \begin{bmatrix}
        0.0000 \\
        0.0301 \\
        0.0505
    \end{bmatrix}, \quad
    \theta^{\mathbb{Q}} =
    \begin{bmatrix}
        0.0000 \\
        0.0633 \\
        0.0766
    \end{bmatrix}.
    \]

    \item \textbf{Correlation matrix of innovations:}
    \[
    \rho =
    \begin{bmatrix}
        1.0000 & -0.6303 & -0.4097 \\
        -0.6303 & 1.0000 & 0.2993 \\
        -0.4097 & 0.2993 & 1.0000
    \end{bmatrix}.
    \]

    \item \textbf{Volatilities (diagonal of }$\Sigma$\textbf{):}
    \[
    \text{diag}(\Sigma) =
    \begin{bmatrix}
        0.0027 \\
        0.0045 \\
        0.0070
    \end{bmatrix}.
    \]

    \item \textbf{Speed of mean-reversion (under $\mathbb{P}$):}
    \[
    \kappa^{\mathbb{P}} =
    \begin{bmatrix}
        0.0075 & 0.0000 & 0.0000 \\
        0.0000 & 0.0288 & -0.0233 \\
        0.0000 & 0.0000 & 0.0354
    \end{bmatrix}.
    \]

    \item \textbf{Speed of mean-reversion (under $\mathbb{Q}$):}
    \[
    \kappa^{\mathbb{Q}} =
    \begin{bmatrix}
        0.0000 & 0.0000 & 0.0000 \\
        0.0000 & 0.0233 & -0.0233 \\
        0.0000 & 0.0000 & 0.0233
    \end{bmatrix}.
    \]

    \item \textbf{Market price of risk vector:}
    \[
    \gamma =
    \begin{bmatrix}
        2.7923 \\
        1.2016 \\
        1.7167
    \end{bmatrix}.
    \]

    \item \textbf{Initial Factor Values:}\footnote{Starting values $X_0$ correspond to filtered values of the factors process on January 31, 2022, which is the last of day of the sample that is used to estimate parameters. See Appendix B.2 of \cite{eghbalzadeh2024evaluation} for such values.}
    \[
    X_0 =
    \begin{bmatrix}
        -0.0312 \\
        0.0384 \\
        0.0688
    \end{bmatrix}.
    \]
\end{itemize}

These values are used as the initial condition for simulating the evolution of the yield curve factors over time.

\section{Swaption Pricing Network \label{sec:SwaptionPricingMetho}}

 There is no closed-form solution for the swaption price provided by Equation~\eqref{eq:swaption_price_T_measure}, and thus the swaption sensitivity $\frac{\partial \mathrm{PS}^{}_{t-1}}{\partial X_{t-1}^{(k)}}$ is not available in closed-form either. To calculate such quantity we adopt a hybrid method that combines Monte Carlo simulation with neural network approximation. 

 Our goal is to approximate the mapping from the factor values and time to maturity to the corresponding swaption price.
 The approach proceeds as follows

\begin{enumerate}
    \item \textbf{Simulate factor paths under the physical measure.}  
    We simulate $N_k$ sample paths of the factor process $(X_t)_{t = 0, 1, \dots, T_\beta}$ under the real-world (physical) measure $\mathbb{P}$. 
    This results in an array of $N_k \times (T_\beta + 1)$ samples.

    \item \textbf{Select $N_k$ samples for $X_0$.}  
    From the simulated paths obtained in Step 1, we sample the $N_k$ initial values $\{_{(i)}X_{0}\}_{i=1}^{N_k}$ randomly.\footnote{Any value with the paths, including non-initial values, can be selected.} These represent diverse and realistic starting conditions for the term structure factors.

    \item \textbf{Randomize swaption maturities.}  
    For each of the $N_k$ samples obtained from step 2, we generate an integer $T_\alpha^{(i)}$ uniformly at random from the set $\{1, 2, \dots, T_\alpha\}$. This assigns a random maturity to each sample of the previous step, reflecting a distribution of swaptions across different maturities.

    \item \textbf{Compute swaption prices using the forward measure.}  
    For each pair $(_{(i)}X_0, T_\alpha^{(i)})$, we use the forward-measure swaption pricing formula from \cite{godin2023pricing}:
    \[
    PS\left[0, \mathcal{T}^{(i)}, K, N \right] 
    = \mathbb{E}^{\mathbb{Q}^{T_\alpha^{(i)}}} \left[\mathcal{P}_{T_\alpha^{(i)}} \,\middle|\, \mathcal{F}_0 \right],
    \]
    where $\mathcal{P}_{T_\alpha^{(i)}}$ denotes the discounted swap payoff at maturity $T_\alpha^{(i)}$, and $\mathcal{T}^{(i)}$ is the associated tenor structure. The expectation is evaluated using Monte Carlo simulation under the corresponding $T_\alpha^{(i)}$-forward measure.
\end{enumerate}

This process yields a training dataset consisting of input-output pairs $\{(_{(i)}X_0, T_\alpha^{(i)}), PS^{(i)}\}$, which is then used to train a neural network that approximates the pricing function, as described in Section~\ref{Sec:KANnet}, and its sensitivities for out-of-sample prediction. For the sensitivities, a finite difference approach is considered.

In summary, the objective is to learn a smooth approximation of the pricing function, thereby enabling fast and accurate generalization for out-of-sample factor inputs during hedging and simulation. This model is then used to evaluate swaption prices and sensitivities in real-time, significantly reducing computational overhead while preserving accuracy. 

The type of neural network considered is a fully connected Kolmogorov–Arnold neural network (KAN). The structure of such network is discussed next.

\subsection{Kolmogorov–Arnold Networks\label{Sec:KANnet}}

Kolmogorov–Arnold Networks (KANs) are a recent neural architecture first introduced in \cite{liu2024kan} and inspired by the Kolmogorov–Arnold representation theorem, which asserts that any multivariate continuous function can be written as a finite sum of compositions of univariate continuous functions.

KANs provide a more interpretable alternative to standard FCNNs.
While traditional FCNNs apply fixed linear transformations followed by nonlinear activation functions (such as ReLU or Sigmoid), KANs take a different approach: they apply learnable univariate functions directly to each input coordinate and then combine the results linearly. This design allows KANs to adaptively learn nonlinear transformations on a per-dimension basis, making them highly expressive and naturally suited for capturing structured dependencies.

In this work, we use a KAN to approximate the swaption pricing function. Each univariate function is parameterized as a linear combination of Radial Basis Functions (RBFs), following the FastKAN framework proposed in \cite{li2024kolmogorovarnold}. These Gaussian RBFs are centered along the input axis, allowing the network to capture local patterns in the data with smooth transitions.
This architecture enhances both flexibility and interpretability.

The structure of the KAN model used in this study is summarized as follows:
\begin{itemize}
    \item Each input pair $(X_\tau, \tau)$, consisting of the yield curve factors at time $\tau  \in \{0, 1, \dots, T_\alpha \}$ and the time-to-maturity, passes through a learnable univariate function composed of RBF kernels.
    \item The outputs of these functions are then linearly combined to produce the network output, which represents the swaption price under predefined parameters.
    \item Several such layers are stacked to form a deep KAN model, optionally including residual connections or normalization layers to improve training dynamics.
\end{itemize}

This structure provides the agent with a faster, more accurate, and interpretable approximation of swaption prices, which is particularly advantageous in settings where no closed-form pricing formula is available. For a detailed comparison highlighting why KAN are preferable to standard FCNN architectures in our setting, see Section~\ref{app:mlp_kan}.

\paragraph{Pricing Network Architecture.}
To approximate the swaption pricing function, we employ a three-layer Kolmogorov–Arnold Network (KAN) with layer widths of 8, 16, and 8, respectively. The model takes as input four features: the three term structure factors generated by the DTAFNS model and the time-to-maturity, expressed in years. The network outputs a single scalar corresponding to the price of a payer swaption.

Training data is generated using Monte Carlo simulations, with data for each swaption of maturity $T^{(i)}_{\alpha}$ being generated under the $T^{(i)}_{\alpha}$-forward risk-neutral measure, using $N_k=20{,}000$ sample paths.
However, each swaption in the dataset has the same tenor and strike, each corresponding to these of the swaption we want to price in subsequent simulation experiments. The network is trained for up to 5{,}000 epochs using the Adam optimizer with a learning rate of 0.001, minimizing the smooth L1 loss (Huber loss) between predicted and simulated swaption prices.

Once trained, the pricing network is used to evaluate swaption prices and compute their gradients with respect to the input factors. We employ the central difference method to compute the gradient, as it exhibits greater stability and accuracy compared to the neural network's gradient obtained through automatic differentiation. The computed sensitivities are then used to determine the hedge ratios in the benchmark rho-hedging strategies.

\subsection{KAN vs. FCNN Performance Comparison} \label{app:mlp_kan}

\begin{figure}[h!]
    \centering
    \includegraphics[width=\textwidth]{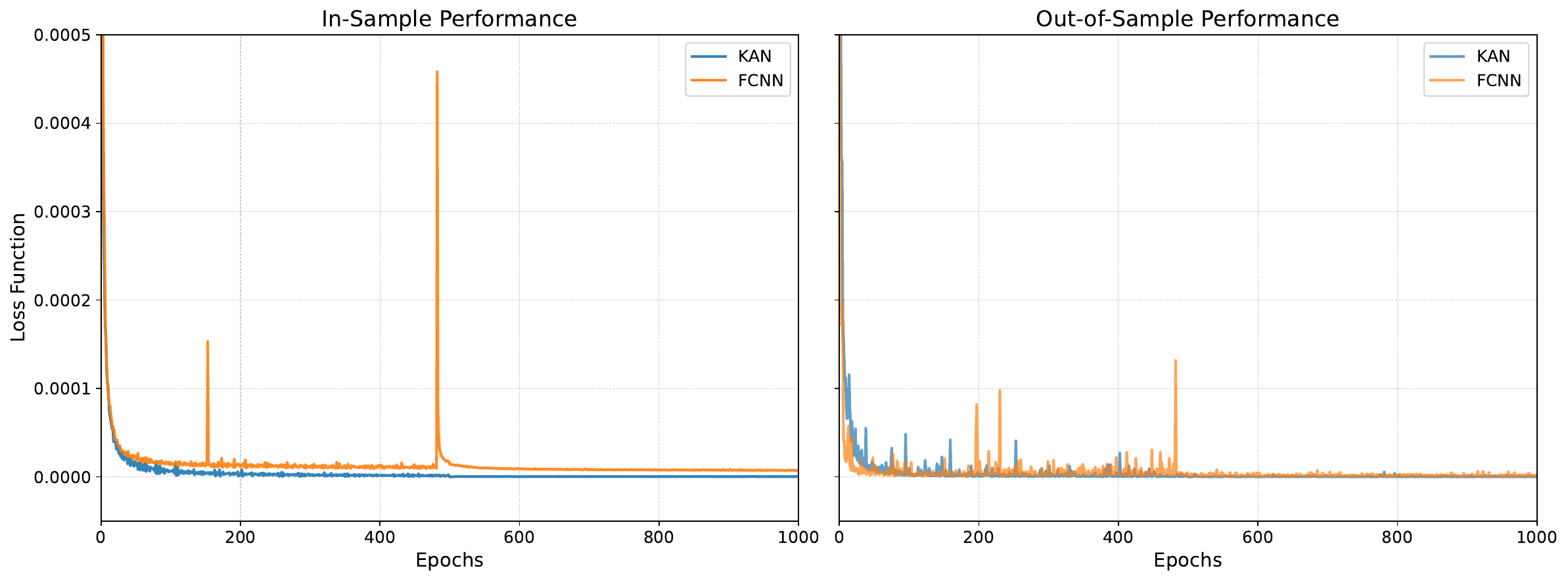}
    \caption{In-sample (left) and out-of-sample (right) performance of KAN and FCNN across 1000 training epochs.}
    \label{fig:performance_curves}
\end{figure}

This section presents a comparative analysis of the Kolmogorov–Arnold Network (KAN) and a standard FCNN for approximating swaption prices. While the KAN architecture consists of 3 layers with widths $[8, 16, 8]$, the FCNN model employs a deeper and wider architecture with 4 layers of respective widths $[512, 1024, 1024, 512]$. Despite its significantly smaller size, the KAN demonstrates superior performance. Indeed, Figure~\ref{fig:performance_curves} illustrates the in-sample (IS) and out-of-sample (OOS) performance trajectories of both models over 1000 training epochs.

The KAN consistently outperforms the FCNN across multiple quantitative metrics. In the in-sample setting, the KAN exhibits a smooth and stable convergence trajectory with minimal variance. In contrast, the FCNN displays several sharp fluctuations, notably around epochs 150 and 500, indicating optimization instability and sensitivity to local minima.
Out-of-sample performance further highlights the KAN's superiority in terms of generalization. The mean OOS error over the 1000 epochs for the KAN is $1.73 \times 10^{-6}$. 
It is substantially lower than that of the FCNN, which is $4.71 \times 10^{-6}$. Additionally, the KAN achieves a significantly smaller generalization gap and reaches stable performance approximately 250 epochs earlier than FCNN (309 vs. 562 epochs). The final OOS error for the KAN remains in the order of $10^{-9}$, an order of magnitude lower than that of the FCNN.
Overall, the KAN achieves:
\begin{itemize}
    \item Lower mean and final IS and OOS errors,
    \item Smaller standard deviations, indicating greater robustness,
    \item Faster convergence to a stable solution,
    \item A tighter generalization gap, confirming enhanced predictive performance.
\end{itemize}

These findings support the conclusion that the KAN offers a faster, more stable, accurate, and generalizable architecture for swaption price approximation compared to conventional feedforward networks, even when using significantly fewer parameters.

\end{document}